\documentclass[print,longbibliography,nofootinbib,superscriptaddress,amsmath,amssymb,aps,noeprints,twocolumn]{revtex4-2}

\usepackage{lineno}
\usepackage{siunitx}
\usepackage{graphicx}
\usepackage{dcolumn}
\usepackage{bm}

\makeatletter
\def\maketitle{
\@author@finish
\title@column\titleblock@produce
\suppressfloats[t]}
\makeatother

\usepackage[english]{babel}

\usepackage{lineno}
\usepackage[colorlinks=true, allcolors=blue]{hyperref}

\begin{document}
\title{Dependence of the extra-cellular diffusion coefficient on the fractions of neurites and cell bodies in gray matter}

\author{Hong-Hsi Lee}
\email[Corresponding author: ]{HLEE84@mgh.harvard.edu}
\affiliation{Athinoula A. Martinos Center for Biomedical Imaging, Department of Radiology, Massachusetts General Hospital,
Charlestown, MA, United States}
\affiliation{Harvard Medical School, Boston, MA, United States}

\author{Ali Abdollahzadeh}
\affiliation{A.I. Virtanen Institute for Molecular Sciences, University of Eastern Finland,
Kuopio, Finland}

\author{Hansol Lee}
\affiliation{Athinoula A. Martinos Center for Biomedical Imaging, Department of Radiology, Massachusetts General Hospital,
Charlestown, MA, United States}
\affiliation{Harvard Medical School, Boston, MA, United States}

\author{Ricardo Coronado-Leija}
\affiliation{Center for Advanced Imaging Innovation and Research (CAI2R), Department of Radiology, New York University School of Medicine, 
New York, NY, United States}

\author{Els Fieremans}
\affiliation{Center for Advanced Imaging Innovation and Research (CAI2R), Department of Radiology, New York University School of Medicine, 
New York, NY, United States}

\author{Susie Y. Huang}
\affiliation{Athinoula A. Martinos Center for Biomedical Imaging, Department of Radiology, Massachusetts General Hospital,
Charlestown, MA, United States}
\affiliation{Harvard Medical School, Boston, MA, United States}

\author{Dmitry S. Novikov}
\affiliation{Center for Advanced Imaging Innovation and Research (CAI2R), Department of Radiology, New York University School of Medicine, 
New York, NY, United States}

\date{\today}

\begin{abstract}
\noindent \textbf{Purpose:} The dependence of the long-time (tortuosity) limit of the extra-cellular diffusivity on the intra-cellular volume fraction is of fundamental importance for microstructure modeling. 
While such dependencies have been explored for the white matter, the tortuosity limit in  gray matter is unknown due to complex cell composition and  geometry.
Here we rationalize and validate numerically the analytical relation between the extra-cellular diffusivity and intra-cellular fractions of cell bodies (somas) and neurites.

\noindent \textbf{Methods:} 
The tortuosity relation for extra-cellular diffusivity qualitatively follows from effective medium theory, coarse-grained by diffusion outside somas (spheres) and neurites (cylinders), respectively. This problem is equivalent to finding the overall conductivity in a medium of grains in a matrix, with methodology dating back to the 19th century.
We extend the effective medium methodology to populations of impermeable spheres and randomly oriented cylinders with various volume fractions, yielding closed-form expressions corroborated by Monte Carlo simulations. 

\noindent \textbf{Results:} 
We establish the power-law scaling of the extra-cellular diffusivity with the volume fractions of the extra-soma and extra-neurite spaces.
We further evaluate the proposed framework using simulations in realistic tissue geometries.

\noindent \textbf{Conclusion:} Theory and simulations relate extra-cellular tortuosity to soma and neurite fractions, potentially offering a  diffusion MRI protocol design optimized for in vivo assessment of soma size and soma/neurite fractions within clinical scan times. Such in vivo measurements can be used to study development, aging and neurodegenerative disorders. 

\end{abstract}

\maketitle
\newpage

\section{Introduction}

\begin{figure*}[t!]
\centering
\includegraphics[width=0.75\textwidth]{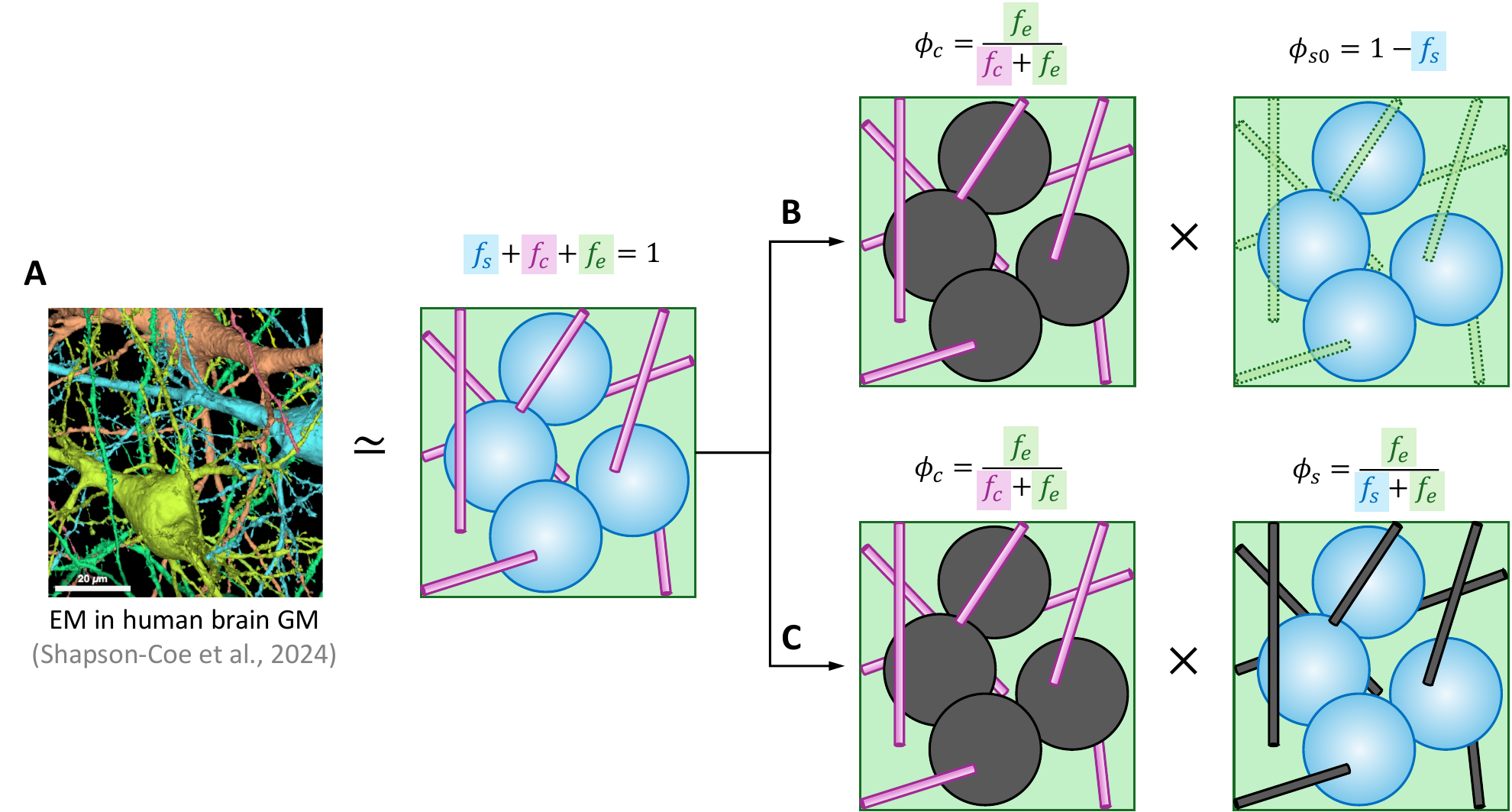}
\caption[]{\textbf{} \textbf{A.} Gray matter in the human brain consists of complex micro-geometries with abundant cell bodies (somas, $>$10 \textmu m in diameter) and highly dispersed thin neurites ($\sim$1 \textmu m in diameter) \cite{shapson2024petavoxel}, which can be considered as a collection of randomly packed spheres (blue) and randomly oriented thin cylinders (magenta) with an extra-cellular space (green). \textbf{B.} Sequential coarse-graining solution: To derive the extra-cellular tortousity of this gray matter model, we consider a two-step diffusion coarse-graining process that has a specific order due to the length scale separation of soma and neurite (Section~\ref{sec:sequential}). \textbf{C.} Simultaneous coarse-graining solution: Alternatively, we propose a two-step diffusion coarse-graining process that has no specific order (Section~\ref{sec:simultaneous}), insensitive to the length scale separation of soma and neurite.} 
\label{fig:phys}
\end{figure*}

Gray matter in the human brain is composed of neuronal cell bodies (somas), dendrites and axons (neurites), along with glial cells and the extracellular space \cite{shapson2024petavoxel}.  
Alterations in soma size and soma/neurite density have been observed in normal aging and neurodegenerative diseases \cite{scott1992amygdala,andrade2013cellad,li2021cellad,lee2024sandi}. 
These findings call for an in vivo histological imaging technique to evaluate soma and neurite microstructure that could be used for earlier detection of neurodegeneration and monitoring of treatment response.

Diffusion MRI (dMRI) is an imaging technique sensitive to tissue micro-geometries at a length scale commensurate with the cell size \cite{novikov2019review}, three orders smaller than the image voxel size, allowing in vivo estimation of tissue histopathology.
Compared with conventional diffusion signal representations, such as diffusion tensor imaging \cite{basser1994dti} and diffusion kurtosis imaging \cite{jensen2005dki}, biophysical modeling of dMRI enables characterization and interpretation of tissue microstructure in terms of specific parameters, such as cell size \cite{alexander2010activeax,assaf2008axcaliber,barazany2009axcaliber,sepehrband2016ghighg,benjamini2016dpfg,fan2020adm,veraart2020highb,palombo2020sandi}, intra- and extra-cellular diffusivities and water fractions \cite{fieremans2011wmti,zhang2012noddi,kaden2016smt,novikov2018rotinv}.

Relating the long-time limit of extra-cellular diffusivity $D$ to packing geometry of impermeable cells, with the extra-cellular volume fraction $\phi$ and intrinsic diffusivity $D_0$, maps onto a classic problem of finding electrical dc conductivity $\sigma(\phi)$ of a mixture made of non-conducting grains with volume fraction $1-\phi$, immersed in a medium with  conductivity $\sigma_0$ \cite{torquato2002emt}.
As the diffusion coefficient is normalized per diffusing molecule, while the conductivity grows with the number of charge carriers, the ratio by which the diffusion coefficient is reduced,
\begin{equation} \label{eq:tortuosity-def}
    \frac{1}{\Lambda(\phi)} \equiv \frac{D(\phi)}{D_0} = 
    \frac{\sigma(\phi)}{\sigma_0}\cdot \frac1{\phi} \,,  
\end{equation}
involves the dimensionless conductivity $\sigma/\sigma_0$ further divided by the extra-grain fraction $\phi$.\cite{latour1994time} 
It is conventional to introduce the dimensionless {\it tortuosity} factor $\Lambda(\phi)\geq 1$ that tells by how much the extra-cellular diffusivity is suppressed relative to free $D_0$.     
As the effective medium theory (EMT) is naturally developed for conductivity, rather than for tortuosity,  in what follows we will consider scalings for $\sigma(\phi)/\sigma_0$ and translate to tortuosity via Eq.~(\ref{eq:tortuosity-def}).

The mapping (\ref{eq:tortuosity-def}) onto conductivity relates the tortuosity problem to a test-bed of effective medium theories starting from Maxwell's work in the 19th century \cite{maxwell1873em,maxwell1904emt,bruggeman1935emt,sen1981emt,hashin1962emt,weissberg1963effective,landauer1978emt,berryman1980long,tomadakis1993transport,latour1994time,torquato2002emt,giordano2003emt,novikov2010emt,novikov2012emt}.
The EMT-derived tortuosity relation works well when the grains are loosely packed with low volume fraction, or when the electrical (or equivalently diffusion) properties of grains and matrix are similar. 
However, neurons and glial cells in the brain are tightly packed with high intra-cellular volume fractions $\sim$ 80\% \cite{pallotto2015ecs}, making it more challenging to derive the accurate tortuosity relation.

In white matter, where axons are tightly packed and effectively impermeable ($\phi\sim$ 0.2–-0.5), traditional effective medium models lose predictive power, with classical bounds \cite{hashin1962emt} and self-consistent schemes \cite{bruggeman1935emt,sen1981emt} erring by as much as 100\% for the diffusion coefficient $D$ transverse to the tracts \cite{novikov2012emt}. 
The difficulty arises since in this 2-dimensional (2d) setting, each axon acts as a strong hindrance to diffusion, reshaping the trajectories of water molecules that are further complicated by the close proximity of neighboring axons, which eliminate any obvious small parameter to guide perturbative analysis. 

Interestingly, the 2d tortuosity relation of densely packed axons (disks) in white matter was solved by accounting for the axon diameter distribution \cite{novikov2012emt}, approximated by a small population for the bulk distribution and another large population for the tail of the distribution.
{The result is notably different from the 2d mean-field tortuosity relation $\Lambda=1/\phi$ \cite{bruggeman1935emt,sen1981emt}.
To achieve a better agreement with simulations in tightly packed geometries,} diffusion coarse-graining was modeled in two stages\cite{novikov2012emt},  by first averaging the effect of small axons over scales larger than their size, and subsequently treating the rare large axons as an additional restriction to the reduced diffusivity.
However, due to the complexity of gray matter microstructure, the tortuosity relation for gray matter --- an essentially 3d problem --- has remained unknown. 

Under a simplified biophysical picture (Figure~\ref{fig:phys}A), gray matter can be considered as a collection of somas (spheres) and neurites (sticks) embedded into an extra-cellular space (isotropic hindered diffusion) \cite{palombo2020sandi}. 
For gray matter modeling, a tortuosity relation of highly aligned (2d) cylinders along with randomly packed spheres has been applied \cite{gyori2021sandi}.
However, this model was originally designed for the bovine optic nerve \cite{stanisz1997simulation}, where the mean-field tortuosity relation did not apply well due to densely packed axons, and the highly aligned cylinders did not match the biological picture of highly dispersed neurites in gray matter \cite{shapson2024petavoxel}.

In this work, to elucidate the tortuosity relation of extra-cellular space in gray matter, we start from a  picture of diffusion coarse-graining in a medium composed of randomly packed spheres (somas) and randomly oriented, randomly positioned cylinders (neurites) using effective medium theory (Figure~\ref{fig:phys}). 
Furthermore, we perform Monte Carlo (MC) simulations of diffusion in the extra-cellular space of media composed of randomly packed spheres and dispersed cylinders, and validate the derived tortuosity relation across varying sphere and cylinder volume fractions, including the extreme case of densely packed geometries. 
We additionally perform simulations in a realistic extra-cellular space geometry segmented from electron microscopy data of the mouse olfactory bulb \cite{pallotto2015ecs}.
The significance of the developed tortuosity relation lies in reducing the data requirements for dMRI modeling in gray matter, thereby facilitating clinical translation.

\section{Theory}
Tissue micro-geometry in brain gray matter is complex \cite{shapson2024petavoxel}, with an abundance of cell bodies and highly dispersed dendrites and axons.
To resolve the tortuosity relation of extra-cellular diffusivity in gray matter, we simplify the gray matter microstructure, representing it as a collection of randomly packed spheres and randomly oriented cylinders. 
The spheres do not overlap with each other, while the cylinders may overlap with each other, mimicking dendritic arborization (branching) \cite{jan2010branching}, spines and leaflets \cite{palombo2018leaflet}.
First, we discuss two simple cases, a medium composed of either spheres or cylinders; then we consider the more complex case of a medium composed of both spheres and cylinders.

\subsection{Differential addition}
\label{sec:add}

The idea of effective-medium conductivity originates from the following 19th century observation: for a small fraction $f\ll 1$ of spheres of conductivity $\sigma_1$ embedded in the matrix with conductivity $\sigma_0$, the  conductivity $\sigma(f)$ of this composite medium follows the following generalized Clausius-Mossotti relation \cite{maxwell1873em,maxwell1904emt} 
\begin{equation} \label{CM}
    \frac{\sigma-\sigma_0}{\sigma+2\sigma_0} 
    = f \cdot \frac{\sigma_1-\sigma_0}{\sigma_1+2\sigma_0} \,.
\end{equation}
(The original relation was formulated for the dielectric permittivity of a mixture, as a solution to the same Laplace equation.)   
Essentially, a large sphere of volume $V$ and conductivity $\sigma$ embedded in a matrix of $\sigma_0$ distorts the current (or is polarized) approximately in the same way as $N$ small spheres (of volume $v_1$) with conductivity $\sigma_1$ and total volume $v = Nv_1$ within the same $V$, where $f = v/V \ll1$; this is a mean-field view that the ``polarizabilities''  $(\sigma_1-\sigma_0)/(\sigma_1+2\sigma_0)$ add up. 

Bruggeman\cite{bruggeman1935emt} was first to realize that Eq.~(\ref{CM}) can be written in a differential form (from a modern perspective, it appears as a precursor to the real-space renormalization group equation): Suppose we have already added spheres with total volume $v$ into a matrix with volume $V-v$, such that the spheres have a finite volume fraction $f=v/V$; 
the whole composite medium has an effective conductivity $\sigma_0\to \sigma(v)$. Adding a bit more spheres of $v\to v+\text{d}v$ would change the conductivity $\sigma\to \sigma + \text{d}\sigma$; according to Eq.~(\ref{CM}): 
\begin{equation} \label{Brugg}
\frac{\text{d}\sigma}{3 \sigma} 
= - \frac12 \, \frac{\text{d}v}{V}\,,
\end{equation}
where we set $\sigma_1\equiv 0$ (non-conducting spheres) in the right-hand side of Eq.~(\ref{CM}) for our case, and used the infinitesimal fraction $\text{d}v/V$ in lieu of $f$.  
Since 
\begin{equation*}
    \text{d}f = \frac{v+\text{d}v}{V+\text{d}v} - \frac{v}{V}
    \simeq \frac{\text{d}v}{V} (1-f)\,,
\end{equation*}
we can write Eq.~(\ref{Brugg}) as 
\begin{equation} \label{spheres}
    \text{d} \ln \sigma = -\frac{3}{2} \frac{\text{d}f}{1-f}
=  \frac{3}{2}\, \text{d} \ln \phi \,, \quad \phi = 1-f \,.
\end{equation}
The solution of this differential equation, with the initial condition $\sigma|_{\phi=1}=\sigma_0$, is $\sigma/\sigma_0 = \phi^{3/2}$,\cite{bruggeman1935emt,sen1981emt} yielding the mean-field 3d tortuosity $\Lambda_s(\phi) = \phi^{-1/2}$ for the spheres, cf. Eq.~(\ref{eq:tortuosity-def}). 
This so-called Archie's law\cite{sen1981emt} generalizes to 2-dimensional disks, for which Eq.~(\ref{CM}) has $\sigma+\sigma_0$ in the denominator instead of $\sigma+2\sigma_0$, such that the corresponding scaling $\sigma/\sigma_0 = \phi^2$, $\Lambda_{\rm disk}(\phi) = 1/\phi$. Generalizing for a mixture of randomly oriented ellipsoids, one obtains\cite{giordano2003emt}
\begin{equation}
    \text{d} \ln \sigma = \nu\,\text{d}\phi \,, 
    \quad \nu = \frac{1}{3} \sum_{j=1}^3 \frac{1}{1-L_j}
\end{equation}
where $L_j$ are the demagnetizing factors.
For the spheres,  all $L_j=1/3$. For randomly oriented cylinders, 
$L_1=L_2 =1/2$, and $L_3 = 0$. This yields $\sigma=\sigma_0 \, \phi^{\nu}$ with $\nu=\nu_{c,s}$ for randomly oriented cylinders ($c$) and spheres ($s$):  
\begin{equation} \label{nunu}
    \nu_s=\frac{3}{2} \,, \quad \nu_c=\frac{5}{3} \,,
\end{equation}
cf. Eq.~(\ref{spheres}), 
such that the tortuosity $\Lambda_c(\phi) = \phi^{-2/3}$ for randomly oriented cylinders. 

\subsection{Mixture of cylinders and spheres}

Imagine now composing a medium of two kinds of grains --- non-conducting cylinders and spheres with fractions $f_{c,s}=v_{c,s}/V$. By the above mean-field logic, we have \cite{markov2014determination}
\begin{equation} \label{add_cs}
    \text{d}\ln\sigma = -\frac{\nu_c\,\text{d}v_c}{V} -\frac{\nu_s\,\text{d}v_s}{V}\,.
\end{equation}
To solve this differential equation for $\sigma(v_c,v_s)$, we need to further specify the order of adding different grains to the matrix of conductivity $\sigma_0$ --- i.e., to provide the relation between $v_c$ and $v_s$. The simplest way is to keep a constant ratio $\lambda=v_c/v_s = f_c/f_s$ at all times; when $\text{d}v_c/\text{d}v_s = v_c/v_s$, 
\begin{equation} \notag
\frac{\text{d} f_{c,s}}{1-f_c-f_s} = \frac{\text{d} v_{c,s}}{V} 
\end{equation}
which  turns the right-hand side of Eq.~(\ref{add_cs}) into a full derivative: 
\begin{equation} \notag
    \text{d}\ln\sigma = -\frac{\nu_c\text{d}f_c}{1-f_c-f_s} -\frac{\nu_s\text{d}f_s}{1-f_c-f_s} =\nu\cdot  \text{d}\ln \phi\,,
\end{equation}
where $\phi$ is the overall extra-cellular volume fraction,
\begin{equation} \label{eq:fe-def}
    \phi=f_e=1-f_c-f_s\,.
\end{equation}
The solution for $\sigma|_{\phi=1}=\sigma_0$ is 
\begin{equation} \label{eq:sigma-emt-diff}
    \frac{\sigma_{\rm mix}}{\sigma_0}= \phi^\nu\,, 
    \quad 
    \frac{D_{\rm mix}}{D_0} = \phi^{\nu-1} \,,
\end{equation}
where the exponent 
\begin{equation} \label{nu}
    \nu=\frac{\lambda}{1+\lambda}\nu_c + \frac{1}{1+\lambda}\nu_s
    = \frac{f_c\, \nu_c + f_s\, \nu_s }{f_c+f_s}
\end{equation}
is the volume-weighted average of the exponents (\ref{nunu}) for both kinds of grains. 
The above solution nicely interpolates between either kind of grains, $f=f_c$ or $f=f_s$ in Section~\ref{sec:add}, via its exponent (\ref{nu}). Experience tells\cite{novikov2012emt} that such mean-field theory can work down to moderate extra-cellular fractions $\phi$, but would eventually break down for dense packings, $\phi\ll 1$. In what follows, we will explore this solution (\ref{eq:sigma-emt-diff})--(\ref{nu}) numerically and identify two other empirical solutions that may perform better for low $\phi$. 

\subsection{Sequential coarse-graining}
\label{sec:sequential}
In the gray matter, soma size ($>$10 \textmu m in diameter) is much larger than neurite thickness ($\sim$1 \textmu m in diameter) \cite{shapson2024petavoxel}. 
To account for the size disparity between soma and neurites, we propose a sequential two-step diffusion coarse-graining process, noting that swapping the order of coarse-graining over different structures changes the results \cite{novikov2012emt} (Figure~\ref{fig:phys}B).

The diffusion length increases with the square root of the time elapsed \cite{einstein1905diffusion}. 
Given a strong separation of scales, at shorter times diffusion coarse-grains the finer objects (neurites/cylinders), yielding an effective medium with a reduced conductivity $\sigma_c=\sigma_0\, \phi_c^{\nu_c}$ in the background of the larger objects (soma/spheres). 
Here the extra-cylinder fraction ({in the space not occupied by spheres}) is given by
\begin{equation} \label{eq:phi-c}
    \phi_c=\frac{f_e}{f_c+f_e}=1-\frac{f_c}{1-f_s} = \frac{\phi}{1-f_s}\,, 
\end{equation}
with $\phi$ in Eq.~(\ref{eq:fe-def}). 
The reduced conductivity $\sigma_c$ is then used as that of a uniform matrix, as an input to the second coarse-graining to account for the hindrance induced by the  spheres/somas at longer times, yielding
\begin{equation} \label{cond_seq}
    \sigma_{\rm seq} = \sigma_c \cdot (1-f_s)^{\nu_s} 
    = \sigma_0 \, \phi_c^{\nu_c} \cdot (1-f_s)^{\nu_s} \,.
\end{equation}
Equivalently, we could start from the 
effective diffusivity 
$D_c/D_0 = (\sigma_c/\sigma_0)/\phi_c = \phi_c^{\nu_c - 1}$
of a uniform matrix, and further correct it by the factor $(1-f_s)^{\nu_s-1}$: 
\begin{equation} \label{eq:tortuosity-sequential}
    \frac{D_{\rm seq}}{D_0} = 
    \left(1-\frac{f_c}{1-f_s}\right)^{2/3} \cdot 
    (1-f_s)^{1/2}\,.
\end{equation}
Note that the conductivity (\ref{cond_seq}) and diffusivity (\ref{eq:tortuosity-sequential}) satisfy Eq.~(\ref{eq:tortuosity-def}), since using Eq.~(\ref{eq:phi-c}) leads to 
\begin{equation*}
    \frac{\sigma_\text{seq}}{\sigma_0}\cdot\frac{1}{\phi}
    =\phi_c^{\nu_c-1} (1-f_s)^{\nu_s-1}
    =\frac{D_\text{seq}}{D_0}\,.
\end{equation*}

As we will see below, this solution empirically works for loosely packed cells (e.g., $f_c+f_s\leq0.3$) with the size disparity between large spheres and thin cylinders \cite{novikov2012emt}. To lift the requirement of size disparity, next we propose an alternative, empirical solution, with simultaneous diffusion coarse-graining over cylinders and spheres, different from Eqs.~(\ref{eq:sigma-emt-diff})--(\ref{nu}). 

\begin{figure*}[t!]
\centering
\includegraphics[width=0.675\textwidth]{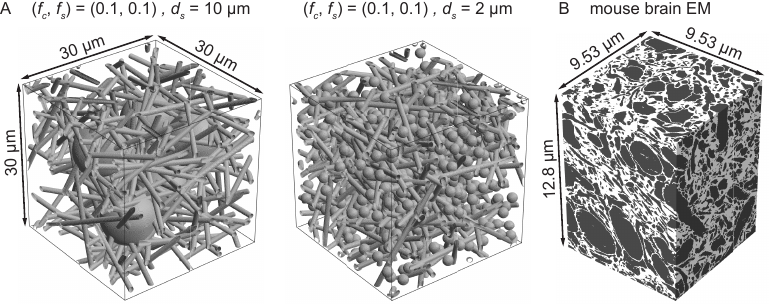}
\caption[]{\textbf{} \textbf{A.} Gray matter-mimicking micro-geometry composed of randomly packed spheres and randomly oriented cylinders with intra-cylindrical and intra-spherical volume fractions ($f_c$, $f_s$) and sphere diameter $d_s$.
\textbf{B.} A 3d SBEM dataset of extra-cellular-space-preserved tissue, sampled from a mouse olfactory bulb \cite{pallotto2015ecs} (Section~\ref{sec:mouse-em}).
} 
\label{fig:geometry}
\end{figure*}

\subsection{Simultaneous coarse-graining}
\label{sec:simultaneous}
Given that the order of coarse-graining processes originates from the size disparity between cylinders and spheres, we hypothesize that a simultaneous coarse-graining process for cylinders and spheres would potentially yields a solution applicable to the case without size disparity. 
In other words, diffusion homogenizes the fine details of the cylinders and spheres on a similar time scale (Figure~\ref{fig:phys}C), {such that neither compartment can be treated as effectively coarse-grained first}. This suggests the following 
guess for the diffusivity of a simultaneously coarse-grained mixture: 
\begin{align} \label{eq:tortuosity-simultaneous} \notag
    \frac{D_{\rm sim}}{D_0} 
    &= \phi_c^{\nu_c-1} \phi_s^{\nu_s-1} \\
    &= \left(1-\frac{f_c}{1-f_s}\right)^{2/3} \cdot \left(1-\frac{f_s}{1-f_c}\right)^{1/2}\,,
\end{align}
where, analogously to Eq.~(\ref{eq:phi-c}), we formally introduced 
the extra-sphere fraction 
\begin{equation} \label{eq:phi-s-A}
    \phi_s = \frac{f_e}{f_s+f_e} = 1-\frac{f_s}{1-f_c}
\end{equation}
in the space not occupied by the cylinders. 
This solution is an empirical guess and does not strictly follow the EMT, since there is no justifiable conductivity that is related to Eq.~(\ref{eq:tortuosity-simultaneous}) via Eq.~(\ref{eq:tortuosity-def}). 
Therefore, it should be viewed as a phenomenological interpolation that captures the combined hindrance effects of cylinders and spheres in regimes where their contributions are not hierarchically separable.
This formulation is expected to be most applicable in densely packed cylinders and spheres with comparable characteristic length scales, while in dilute limits with the presence of strong size disparity, the sequential coarse-graining solution (\ref{eq:tortuosity-sequential}) remains more physically justified.
Surprisingly, despite its heuristic nature, we find that Eq.~(\ref{eq:tortuosity-simultaneous}) provides accurate predictions across a broad range of simulated configurations as shown below.

\section{Methods}

\subsection{Micro-geometry generation}
\label{sec:geometry}
Our goal was to generate a gray matter–mimicking micro-geometry composed of spheres and cylinders, representing cellular components such as somas and neurites, respectively. The simulation domain was a cube of side length $L$, within which we sequentially placed two populations of structures: non-overlapping, randomly packed spheres and randomly oriented, partially overlapping cylinders. The process and associated volume fraction definitions were detailed below.

First, $N_s$ spheres of identical diameter $d_s=2r_s$ were placed, assuming a random close-packing scheme where no two spheres overlapped, and all spheres were entirely contained within the cube \cite{donev2005packing}. The resulting packing achieved a target intra-spherical volume fraction $f_s$, defined as
\begin{equation} \notag
    f_s = \frac{V_s}{L^3}\,.
\end{equation}
where $V_s=N_s\cdot \tfrac{4\pi}{3}r_s^3$ is the total volume of all non-overlapping spheres, including any overlapping volume between spheres and cylinders. For the given sphere radius, cube size, and intra-spherical volume fraction, we calculated the number of spheres accordingly.

After spheres were placed, $N_c$ cylinders of identical diameter $d_c=2r_c$ were added to the geometry. Each cylinder was generated using a homogeneous Poisson line process (mean free path randomness)\cite{coleman1969random} to ensure uniform sampling of orientations and positions:
\begin{enumerate}
  \item We defined a spherical domain of radius $R=5L$, centered at the cube's center.
  \item A direction vector $\mathbf{n}$ was sampled uniformly over a unit sphere.
  \item A random point $\mathbf{x}$ was selected within the spherical domain, and a line segment was constructed from the point of intersection on one side of the spherical domain to the opposite side, resulting in a finite-length segment enclosed in the spherical domain.
  \item Each line was checked for intersection with the cube. Only lines that intersected the cube were retained.
  \item The segment of each intersecting line contained within the cube was used as the cylinder axis.
  \item A solid cylinder of fixed diameter $d_c$ was constructed along each line segment.
\end{enumerate}
To achieve the target intra-cylindrical volume fraction $f_c$, the number of cylinders was determined by \cite{chiu2013stochastic} (Supplementary Section~\ref{sec:app})
\begin{equation} \label{eq:N-cyl}
    N_c = -\frac{L^3}{\langle v_c\rangle} \cdot \ln\left(1 - \frac{f_c}{1 - f_s} \right)\,,
\end{equation}
where 
\begin{equation} \label{eq:v-cyl}
    \langle v_c\rangle=\tfrac{2}{3}\pi r_c^2L
\end{equation}
was the average cylinder volume inside the cube.
This approach ensured that the cylinders collectively occupied the correct fraction of the free volume not already filled by spheres, while preserving realistic microstructural overlap and orientation statistics.

We created multiple micro-geometries composed of identical cylinders of 1 \textmu m in diameter with varying volume fractions $f_c=$ 0, 0.1, 0.2, 0.3, 0.4, 0.5 and identical spheres of either 10 \textmu m or 2 \textmu m in diameter with varying volume fractions $f_s=$ 0, 0.1, 0.2, 0.3, 0.4, 0.5, leading to 46 different micro-geometries (Figure~\ref{fig:geometry}A), excluding the cases of $f_c=f_s=0$ and $f_c=f_s=0.5$.

After generating micro-geometries in a 3-dimensional continuous space enclosed within the cube of side length $L=$ 30 \textmu m, we pixelated each geometry into a 600\texttimes600\texttimes600 matrix with an isotropic pixel size of 50 nm. The $f_c$ and $f_s$ were calculated again by counting the number of pixels in each compartment. The relative error of $f_c$ ($f_s$) was less than 3.5\% (0.005\%).
{The relative error of sphere diameter was about 1\% or less.}

\subsection{Mouse electron microscopy}
\label{sec:mouse-em}
To demonstrate our theory in realistic extra-cellular space of brain gray matter, we analyzed a publicly available three-dimensional serial block-face scanning electron microscopy (SBEM) dataset of extra-cellular-space-preserved tissue, sampled from a mouse olfactory bulb (Figure~\ref{fig:geometry}B) \cite{pallotto2015ecs}.
The original sample had a volume of 10\texttimes10\texttimes12.8 \textmu m\textsuperscript{3} with a voxel resolution of 9.8\texttimes9.8\texttimes25 nm\textsuperscript{3}.
We segmented all neurites 
in the sample using DeepACSON \cite{abdollahzadeh_cylindrical_2021, abdollahzadeh2021deepacson}, followed by instance segmentation initiated from the neurite skeletons provided in the original dataset. We considered the space outside segmented neurites as the extra-cellular space.

The binary mask of the extra-cellular space was smoothed by applying a signed distance transform, Gaussian smoothing with a width of 0.72 voxel, and thresholding at 0.
We cropped the edge of each slice by 250 nm, where segmentation was relatively challenging and less precise, resulting in a final volume of 9.53\texttimes9.53\texttimes12.8 \textmu m\textsuperscript{3}.
For diffusion simulations, we downsampled the extra-cellular binary mask to an isotropic voxel resolution of 25\texttimes25\texttimes25 nm\textsuperscript{3}.
The extra-cellular fraction was $f_e=$ 0.353.
The neurites were cylinder-like structures and had a fraction of $f_c=$ 0.647.
The sample contained no cell bodies \cite{pallotto2015ecs}, such that sphere-like structures had a fraction $f_s=$ 0.

\subsection{Monte Carlo simulations}
We performed Monte Carlo (MC) simulations in (i) pixelated micro-geometries composed of impermeable spheres and cylinders, created in Section~\ref{sec:geometry}, {and (ii) a binary mask of realistic extra-cellular space from mouse SBEM (Section~\ref{sec:mouse-em})} using the RMS toolbox \cite{fieremans2018cookbook,lee2020axial,lee2020radial,lee2021rms} implemented in CUDA C++ for diffusion in extra-cellular space.
In each simulation, we applied 1\texttimes10\textsuperscript{5} random walkers, each diffusing with a step size of $\delta x=$ 45 nm {(gray matter-mimicking substrates) and 22.5 nm (mouse SBEM)}, a time step of $\delta t=\delta x^2/(6D_0)$, and an intrinsic diffusivity $D_0=$ 1, 2, or 3 \textmu m\textsuperscript{2}/ms in extra-cellular space. 
We calculated the apparent diffusivity based on the displacement cumulant and averaged them over 100 directions to calculate the mean diffusivity (MD) at diffusion times $t=$ 1--100 ms.
To estimate the extra-cellular diffusivity $D$ in the long time limit, we fitted the following functional form to $\text{MD}(t)$ to factor out the diffusivity time-dependence \cite{novikov2014meso}:
\begin{equation} \label{tail}
    \text{MD}(t)=D + \frac{c}{t}\,,
\end{equation}
where $c$ is the strength of hindrance due to the structural disorder. (Note that, technically, one should utilize the $\ln(t/t_c)/t$ tail in Eq.~(\ref{tail}) arising from random-rods disorder with correlation time $t_c$ \cite{novikov2014meso}. 
However, within the time range considered here, the simpler $1/t$ form yielded a numerically similar tortuosity limit $D$ due to the slowness of $\ln t$, while also providing a more robust fit with fewer free parameters.)
We calculated the extra-cellular tortuosity (\ref{eq:tortuosity-def}) using the simulated $D$ and compared its values with the mixture solution (\ref{eq:sigma-emt-diff})--(\ref{nu}), sequential (\ref{eq:tortuosity-sequential}) and simultaneous coarse-graining solutions (\ref{eq:tortuosity-simultaneous}).

To evaluate the effect of diffusivity time-dependence of wide pulse-gradient sequences, we calculated diffusion signals based on diffusional phase in each simulation at four $b$-values $b=$ 0.5, 1, 1.5, 2 ms/\textmu m\textsuperscript{2} in 20 directions per $b$-shell with inter-pulse interval $\Delta$ and pulse width $\delta$ at ($\Delta$, $\delta$) = (13, 6) ms, (19, 8) ms, and (43.4, 13.7) ms, corresponding to protocols on Connectome 2.0 scanner (500 mT/m maximum gradient strength) \cite{ramos2025connectome2}, Connectome 1.0 scanner (300 mT/m) \cite{tian2022comprehensive}, and clinical scanner (HCP Lifespan, 80 mT/m) \cite{harms2018extending,bookheimer2019lifespan}.
We computed the extra-cellular mean diffusivity $\text{MD}(\Delta,\delta)$ from simulated signals and compared its values with the simultaneous coarse-graining solution (\ref{eq:tortuosity-simultaneous}).

Furthermore, we evaluated the effect of diffusivity frequency-dependence of trapezoidal-cosine oscillating-gradient sequences (Supplementary Section~\ref{sec:og-waveform}, Supplementary Figure~\ref{fig:og-waveform}) \cite{van2014vivo,baron2014ogse,xu2020breast,dai2023og,sung2026complementary}. 
We calculated diffusion signals based on diffusional phase in each simulation at four $b$-values $b=$ 0.25, 0.5, 0.75, 1 ms/\textmu m\textsuperscript{2} in 20 directions per $b$-shell with total waveform time of 92.3 ms, mixing time of 10.5 ms, rise time of 0.9 ms, number of cycles over each side of refocusing pulse $N=$ 1 and 2 for nominal frequency $\omega/2\pi=$ 25 and 50 Hz, respectively, corresponding to the IMPULSED protocol on clinical scanners (80 mT/m) \cite{xu2020breast}.
Similarly, we computed the extra-cellular mean diffusivity $\text{MD}(\omega)$ from simulated signals and compared its values with the simultaneous coarse-graining solution (\ref{eq:tortuosity-simultaneous}).
Note that the IMPULSED protocol refers to the biophysical model for cell size imaging \cite{xu2020breast}, rather than the high-gradient IMPULSE system \cite{feinberg2023impulse}.

\begin{figure}[t!]
\centering
\includegraphics[width=0.45\textwidth]{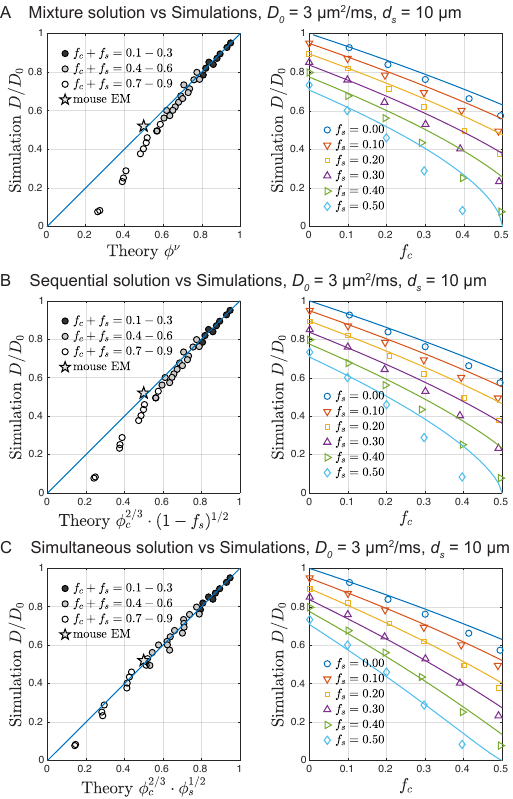}
\caption[]{\textbf{} Comparison of \textbf{A.} mixture solution (\ref{eq:sigma-emt-diff})--(\ref{nu}), \textbf{B.} sequential (\ref{eq:tortuosity-sequential}) and \textbf{C.} simultaneous coarse-graining solutions (\ref{eq:tortuosity-simultaneous}) for extra-cellular tortuosity with simulation results of intrinsic diffusivity $D_0=$ 3 \textmu m\textsuperscript{2}/ms in a medium composed of spheres of $d_s=$ 10 \textmu m in diameter and cylinders of 1 \textmu m in diameter. 
In left panels, simulation results in extra-cellular space of mouse SBEM are shown in pentagrams.
In right panels, simulation results are shown as data points, and theoretical predictions are shown as curves. } 
\label{fig:geometry-D3-ds10}
\end{figure}

\begin{figure*}[t!]
\centering
\includegraphics[width=0.675\textwidth]{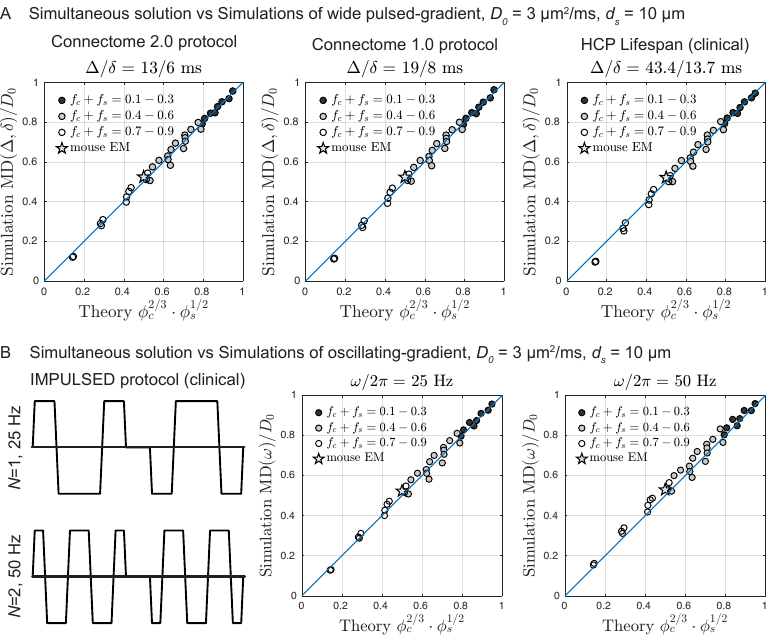}
\caption[]{
\textbf{} Comparison of simultaneous coarse-graining solution (\ref{eq:tortuosity-simultaneous}) for extra-cellular tortuosity with simulation results of \textbf{A.} wide pulsed-gradient and \textbf{B.} oscillating-gradient sequences. 
We performed diffusion simulations with intrinsic diffusivity $D_0=$ 3 \textmu m\textsuperscript{2}/ms in a medium composed of spheres of $d_s=$ 10 \textmu m in diameter and cylinders of 1 \textmu m in diameter. 
For pulsed-gradient waveform, we simulated the time-dependent mean diffusivity $\text{MD}(\Delta,\delta)$ for diffusion protocols (inter-pulse interval $\Delta$, pulse width $\delta$) on three different scanners, such as Connectome 2.0 (500 mT/m) \cite{ramos2025connectome2}, Connectome 1.0 (300 mT/m) \cite{tian2022comprehensive}, and clinical scanners (80 mT/m) \cite{harms2018extending,bookheimer2019lifespan}.
For oscillating-gradient waveform, we simulated the frequency-dependent mean diffusivity $\text{MD}(\omega)$ for the IMPULSED protocol on clinical scanners (80 mT/m) \cite{xu2020breast}.
Here we are showing effective gradient waveforms of oscillating-gradient sequences, where the gradient polarity after the 180$^\circ$ pulse has been inverted to account for the refocusing effect.
} 
\label{fig:pgse-D3-ds10}
\end{figure*}

\section{Results}
\subsection{Monte Carlo simulations}
In diffusion simulations (Figure~\ref{fig:geometry-D3-ds10}, Supplementary Figures~\ref{fig:geometry-D1-ds10}-\ref{fig:geometry-D3-ds2}), the mixture solution (\ref{eq:sigma-emt-diff})--(\ref{nu}) and sequential coarse-graining solution (\ref{eq:tortuosity-sequential}) predicted the simulated extra-cellular tortuosity (\ref{eq:tortuosity-def}) well when the overall intra-cellular volume fraction was low ($f_c+f_s\leq0.3$), whereas the theory deviated from simulations for high intra-cellular volume fractions ($f_c+f_s>0.6$).
In contrast, the simultaneous coarse-graining solution (\ref{eq:tortuosity-simultaneous}) predicted the simulated extra-cellular tortuosity (\ref{eq:tortuosity-def}) well not just for low intra-cellular volume fractions but also the high ones.
This observation was independent of the intrinsic diffusivity $D_0=$ 1--3 \textmu m\textsuperscript{2}/ms and sphere diameters $d_s=$ 10 and 2 \textmu m tested in the simulations.
In the realistic extra-cellular space derived from mouse SBEM data, all solutions performed well.

For wide pulsed-gradient and oscillating-gradient waveforms (Figure~\ref{fig:pgse-D3-ds10}, Supplementary Figures~\ref{fig:pgse-D1-ds10}--\ref{fig:pgse-D3-ds2}), the simultaneous coarse-graining solution (\ref{eq:tortuosity-simultaneous})  predicted the extra-cellular mean diffusivity $\text{MD}(\Delta,\delta)$ (for pulsed-gradient) and $\text{MD}(\omega)$ (for oscillating-gradient) quite well across low to high intra-cellular volume fractions.
For both wide pulsed-gradient and oscillating-gradient waveforms, we observed slightly higher deviations between the proposed solution (\ref{eq:tortuosity-simultaneous}) and simulations with lower intrinsic diffusivity and larger sphere diameter (Figure~\ref{fig:pgse-D1-ds10}).
In addition, the solution performed well in the realistic extra-cellular space derived from mouse SBEM data.
The simulated diffusivity time-dependence across the three diffusion protocols with varying ($\Delta$, $\delta$) was less than 2\%; the simulated diffusivity frequency-dependence across the two frequencies $\omega$ was less than 1.5\%.

\section{Discussion}
In this study, we proposed effective medium theory solutions for the extra-cellular tortuosity relation in a gray matter-mimicking medium consisting of densely packed spheres and randomly oriented cylinders, including a mixture (\ref{eq:sigma-emt-diff})--(\ref{nu}), 
sequential coarse-graining (\ref{cond_seq})--(\ref{eq:tortuosity-sequential}), and simultaneous coarse-graining (\ref{eq:tortuosity-simultaneous}),
and validated the theory via Monte Carlo simulations of diffusion.
The validated theory will be applied to reduce the number of independent fit parameters in biophysical modeling of dMRI and microstructural MRI in brain gray matter \cite{palombo2020sandi,ianucs2022sandimouse}, and may facilitate protocol optimization to shorten scan time on clinical MRI scanners for clinical translation \cite{gyori2021sandi,schiavi2023sandi} and on preclinical scanners for microstructural imaging at high spatial resolution \cite{foo2020magnus,feinberg2023impulse,ramos2025connectome2,wang2019echo}.

The current framework is based on EMT, which assumes weak spatial correlations and homogeneous mixing of grains in a medium. 
In tissues with densely packed cells, spatial correlations and the connectivity of the extra-cellular space may play an important role, since transport becomes increasingly governed by constrained pathways.
Such correlations are not explicitly considered in classical EMT and may lead to deviations, especially at high intra-cellular fractions.
However, we observe that the proposed simultaneous coarse-graining solution (\ref{eq:tortuosity-simultaneous}) remains accurate across densely packed geometries. 
This suggests that, though empirical, the proposed theory may effectively capture the combined influence of volume exclusion and connectivity constraints.

We also note a generally better agreement of the mean-field-like approach of the present 3d setup with MC simulations than in the previously considered 2d problem\cite{novikov2012emt}, which required modeling transport via narrow gorges in-between disks, as well as employing models with the periodic geometry for the packed smaller disks, to reach an agreement with the simulations. Better performance of mean-field approaches is generally expected in higher spatial dimensions, as the role of fluctuations decreases.

Future work will explore the applicability and generalization of these relations to more complex geometries, such as realistic cellular structures observed on microscopic imaging of {human} brain tissue specimens prepared to preserve the extra-cellular space \cite{pallotto2015ecs}, as well as the interconnected extracellular network spanned by tunnels around synapses, and axons and sheets around astrocytes \cite{kinney2013extracellular}. 

Following limitations or simplifying assumptions should be acknowledged. 
Our validation mainly focused on diffusion simulations of the extra-cellular diffusivity in the long-time limit, which remains relevant to many practical scenarios. 
In this limit, the measured diffusivity becomes independent of the specific gradient waveform \cite{novikov2019review}, and clinical scanners typically operate with long effective diffusion times of $\sim$30 ms or longer.
In contrast, shorter diffusion times are more accessible on preclinical systems, high-performance systems \cite{foo2020magnus,huang2021connectome,ramos2025connectome2}, and/or employing oscillating-gradient waveforms \cite{zhu2025engineering}, where residual time- or frequency-dependence, although small, may introduce biases.
Extending our current framework to numerically model these effects across scanner platforms and waveform protocols will be investigated in future work.

Furthermore, our theoretical derivations ignored the effect of structural disorder caused by neurite {branching}, beading, undulation (micro-orientational dispersion), {and glial processes} on the diffusivity tortuosity in the extra-cellular space. 
These microstructural features may affect not only the intra-neurite diffusivity and its time-dependence \cite{novikov2010emt,budde2010bead,novikov2014meso,fieremans2016invivo,lee2018rd,lee2020axial,lee2020gray,lee2024effects,abdollahzadeh2025scattering} but also the coarse-grained extra-cellular diffusivity at long times \cite{coronado2022time,nguyen2026caterpillar}.
The proposed extra-cellular tortuosity relation may change in pathological conditions if the tissue microstructure is altered sufficiently strongly, such that our EMT approach fails.  
The present validation is based on idealized substrates and a single realistic micro-geometry, and extending the analysis to a broader range of realistic substrates is an important direction for future work.

Finally, we assumed impermeable cell membranes and neglected the transcytolemmal water exchange \cite{szafer1995simulation,novikov2011rpbm}. 
In biological tissue, membrane permeability is heterogeneous and cell-type dependent \cite{nagelhus2013physiological}, which may lead to deviations from the proposed tortuosity relation.
The impermeable assumption is expected to be appropriate when the characteristic exchange time is long compared to the diffusion time, such that water molecules predominantly remain within their compartments during the encoding period.
In contrast, for faster exchange regimes or for encoding schemes with enhanced exchange sensitivity \cite{olesen2022smex,jelescu2022nexi,uhl2024nexi,chan2025nexi,chakwizira2023exchange}, deviations from the proposed relations are expected.
Incorporating exchange within the present framework, e.g., via frequency-domain solution \cite{novikov2023exchange} of the Kärger exchange model \cite{fieremans2010karger,olesen2022smex,jelescu2022nexi,uhl2024nexi,chan2025nexi}, may account for these effects and their interplay with structural disorder, and will be examined in future studies.

\section{Conclusions}
We have proposed and validated analytical solutions: a mixture, Eqs.~(\ref{eq:sigma-emt-diff})--(\ref{nu}); sequential coarse-graining (\ref{cond_seq})--(\ref{eq:tortuosity-sequential}); and simultaneous coarse-graining (\ref{eq:tortuosity-simultaneous}), for extra-cellular tortuosity of a medium composed of randomly packed spheres and randomly oriented cylinders, mimicking microgeometry of the gray matter, and validated the theory in Monte Carlo simulations of diffusion with a wide range of tissue parameters.
Surprisingly, the least justifiable simultaneous coarse-graining solution  (\ref{eq:tortuosity-simultaneous}) performs the best. 
The proposed analytical approach helps simplify biophysical models of dMRI in gray matter, facilitating diffusion protocol optimization on clinical scanners and the broader translation of microstructure imaging to the neurological and psychiatric disorders.

\section*{Acknowledgments}

This study was supported by the Office of the Director and National Institute Of Dental and Craniofacial Research of NIH under award number DP5OD031854, by the National Institute of Neurological Disorders and Stroke of NIH under award numbers R01NS118187, R21NS081230, R01NS088040, and U24NS137077, by the National Institute on Aging under award number R21AG085795, and by the National Institute of Biomedical Imaging and Bioengineering of NIH under award numbers P41EB015896, P41EB030006, U01EB026996, UG3EB034875, and P41EB017183.
H.L. was supported by the Bio\&Medical Technology Development Program of the National Research Foundation of Korea under grant number RS-2024-00411768.
A.A. was supported by the Academy of Finland under grant number 360360.


\section*{Data availability statement}
The SBEM dataset was obtained from the publicly available \href{https://datadryad.org/dataset/doi:10.5061/dryad.36h28}{repository} associated with Pallotto et al. \cite{pallotto2015ecs}.

The source code of Monte Carlo simulation of diffusion will be released on our \href{https://github.com/Connectome20}{Github} page.




\bibliography{manuscript_revtex}

@PREAMBLE{
 "\providecommand{\noopsort}[1]{}" 
 # "\providecommand{\singleletter}[1]{#1}%" 
}

@article{alexander2010activeax,
	Author = {Alexander, Daniel C and Hubbard, Penny L and Hall, Matt G and Moore, Elizabeth A and Ptito, Maurice and Parker, Geoff JM and Dyrby, Tim B},
	Journal = {NeuroImage},
	Number = {4},
	Pages = {1374--1389},
	Publisher = {Elsevier},
	Title = {Orientationally invariant indices of axon diameter and density from diffusion {MRI}},
	Volume = {52},
	Year = {2010}}

@article{assaf2008axcaliber,
	Author = {Assaf, Yaniv and Blumenfeld-Katzir, Tamar and Yovel, Yossi and Basser, Peter J},
	Journal = {Magnetic Resonance in Medicine: An Official Journal of the International Society for Magnetic Resonance in Medicine},
	Number = {6},
	Pages = {1347--1354},
	Publisher = {Wiley Online Library},
	Title = {{AxCaliber}: a method for measuring axon diameter distribution from diffusion {MRI}},
	Volume = {59},
	Year = {2008}}

@article{barazany2009axcaliber,
	Author = {Barazany, Daniel and Basser, Peter J and Assaf, Yaniv},
	Journal = {Brain},
	Number = {5},
	Pages = {1210--1220},
	Publisher = {Oxford University Press},
	Title = {In vivo measurement of axon diameter distribution in the corpus callosum of rat brain},
	Volume = {132},
	Year = {2009}}

@article{basser1994dti,
	Author = {Basser, Peter J and Mattiello, James and LeBihan, Denis},
	Journal = {Biophysical journal},
	Number = {1},
	Pages = {259--267},
	Publisher = {Elsevier},
	Title = {{MR} diffusion tensor spectroscopy and imaging},
	Volume = {66},
	Year = {1994}}

@article{budde2010bead,
	Author = {Budde, Matthew D and Frank, Joseph A},
	Journal = {Proceedings of the National Academy of Sciences},
	Number = {32},
	Pages = {14472--14477},
	Publisher = {National Acad Sciences},
	Title = {Neurite beading is sufficient to decrease the apparent diffusion coefficient after ischemic stroke},
	Volume = {107},
	Year = {2010}}

@article{fieremans2016invivo,
	Author = {Fieremans, Els and Burcaw, Lauren M and Lee, Hong-Hsi and Lemberskiy, Gregory and Veraart, Jelle and Novikov, Dmitry S},
	Journal = {NeuroImage},
	Pages = {414--427},
	Publisher = {Elsevier},
	Title = {In vivo observation and biophysical interpretation of time-dependent diffusion in human white matter},
	Volume = {129},
	Year = {2016}}

@article{fieremans2018cookbook,
	Author = {Fieremans, Els and Lee, Hong-Hsi},
	Journal = {NeuroImage},
	Pages = {39--61},
	Publisher = {Elsevier},
	Title = {Physical and numerical phantoms for the validation of brain microstructural {MRI}: {A} cookbook},
	Volume = {182},
	Year = {2018}}

@article{fieremans2010karger,
	Author = {Fieremans, Els and Novikov, Dmitry S and Jensen, Jens H and Helpern, Joseph A},
	Journal = {NMR in Biomedicine},
	Number = {7},
	Pages = {711--724},
	Publisher = {Wiley Online Library},
	Title = {{Monte Carlo} study of a two-compartment exchange model of diffusion},
	Volume = {23},
	Year = {2010}}

@article{jensen2005dki,
	Author = {Jensen, Jens H and Helpern, Joseph A and Ramani, Anita and Lu, Hanzhang and Kaczynski, Kyle},
	Journal = {Magnetic Resonance in Medicine},
	Number = {6},
	Pages = {1432--1440},
	Publisher = {Wiley Online Library},
	Title = {Diffusional kurtosis imaging: the quantification of non-gaussian water diffusion by means of magnetic resonance imaging},
	Volume = {53},
	Year = {2005}}

@article{lee2018rd,
	Author = {Lee, Hong-Hsi and Fieremans, Els and Novikov, Dmitry S},
	Journal = {NeuroImage},
	Pages = {500--510},
	Publisher = {Elsevier},
	Title = {What dominates the time dependence of diffusion transverse to axons: Intra-or extra-axonal water?},
	Volume = {182},
	Year = {2018}}

@article{novikov2014meso,
	Author = {Dmitry S. {Novikov} and Jens H. {Jensen} and Joseph A. {Helpern} and Els {Fieremans}},
	Journal = {Proceedings of the National Academy of Sciences of the United States of America},
	Number = {14},
	Pages = {5088--5093},
	Title = {Revealing mesoscopic structural universality with diffusion},
	Volume = {111},
	Year = {2014}}

@article{novikov2011rpbm,
	Author = {Dmitry S. {Novikov} and Els {Fieremans} and Jens H. {Jensen} and Joseph A. {Helpern}},
	Journal = {Nature Physics},
	Number = {6},
	Pages = {508--514},
	Title = {Random walk with barriers.},
	Volume = {7},
	Year = {2011}}

@article{novikov2010emt,
	Author = {Dmitry S. {Novikov} and Valerij G. {Kiselev}},
	Journal = {NMR in Biomedicine},
	Number = {7},
	Pages = {682--697},
	Title = {Effective medium theory of a diffusion-weighted signal},
	Volume = {23},
	Year = {2010}}

@article{novikov2019review,
	Author = {Dmitry S. {Novikov} and Els {Fieremans} and Sune N. {Jespersen} and Valerij G. {Kiselev}},
	Journal = {NMR in Biomedicine},
	Pages = {e3998},
	Title = {Quantifying brain microstructure with diffusion {MRI}: Theory and parameter estimation},
	Volume = {32},
	Year = {2019}}

@article{novikov2018rotinv,
	Author = {Dmitry S. {Novikov} and Jelle {Veraart} and Ileana O. {Jelescu} and Els {Fieremans}},
	Journal = {NeuroImage},
	Pages = {518--538},
	Title = {Rotationally-invariant mapping of scalar and orientational metrics of neuronal microstructure with diffusion {MRI}},
	Volume = {174},
	Year = {2018}}

@article{zhang2012noddi,
	Author = {Hui {Zhang} and Torben {Schneider} and Claudia A. M. {Wheeler-Kingshott} and Daniel C. {Alexander}},
	Journal = {NeuroImage},
	Number = {4},
	Pages = {1000--1016},
	Title = {{NODDI}: practical in vivo neurite orientation dispersion and density imaging of the human brain.},
	Volume = {61},
	Year = {2012}}

@article{szafer1995simulation,
	Author = {Aaron {Szafer} and Jianhui {Zhong} and John C. {Gore}},
	Journal = {Magnetic Resonance in Medicine},
	Number = {5},
	Pages = {697--712},
	Title = {Theoretical model for water diffusion in tissues.},
	Volume = {33},
	Year = {1995}}

@article{stanisz1997simulation,
	Author = {Greg J. {Stanisz} and Graham A. {Wright} and R. Mark {Henkelman} and Aaron {Szafer}},
	Journal = {Magnetic Resonance in Medicine},
	Number = {1},
	Pages = {103--111},
	Title = {An analytical model of restricted diffusion in bovine optic nerve},
	Volume = {37},
	Year = {1997}}

@article{palombo2018leaflet,
  title={Can we detect the effect of spines and leaflets on the diffusion of brain intracellular metabolites?},
  author={Palombo, Marco and Ligneul, Clemence and Hernandez-Garzon, Edwin and Valette, Julien},
  journal={NeuroImage},
  volume={182},
  pages={283--293},
  year={2018},
  publisher={Elsevier}
}

@article{sepehrband2016ghighg,
	Author = {Sepehrband, Farshid and Alexander, Daniel C and Kurniawan, Nyoman D and Reutens, David C and Yang, Zhengyi},
	Journal = {NMR in Biomedicine},
	Number = {3},
	Pages = {293--308},
	Publisher = {Wiley Online Library},
	Title = {Towards higher sensitivity and stability of axon diameter estimation with diffusion-weighted {MRI}},
	Volume = {29},
	Year = {2016}}

@article{baron2014ogse,
	Author = {Baron, Corey A and Beaulieu, Christian},
	Journal = {Magnetic Resonance in Medicine},
	Number = {3},
	Pages = {726--736},
	Publisher = {Wiley Online Library},
	Title = {Oscillating gradient spin-echo ({OGSE}) diffusion tensor imaging of the human brain},
	Volume = {72},
	Year = {2014}}

@article{abdollahzadeh_cylindrical_2021,
	title = {Cylindrical {Shape} {Decomposition} for {3D} {Segmentation} of {Tubular} {Objects}},
	volume = {9},
	issn = {2169-3536},
	url = {https://ieeexplore.ieee.org/document/9345673/},
	doi = {10.1109/ACCESS.2021.3056958},
	journal = {IEEE Access},
	author = {Abdollahzadeh, Ali and Sierra, Alejandra and Tohka, Jussi},
	month = nov,
	year = {2021},
	pages = {23979--23995},
}

@article{lee2020axial,
  title={A time-dependent diffusion {MRI} signature of axon caliber variations and beading},
  author={Lee, Hong-Hsi and Papaioannou, Antonios and Kim, Sung-Lyoung and Novikov, Dmitry S and Fieremans, Els},
  journal={Communications biology},
  volume={3},
  number={1},
  pages={1--13},
  year={2020},
  publisher={Nature Publishing Group}
}

@article{einstein1905diffusion,
  title={On the motion of small particles suspended in liquids at rest required by the molecular-kinetic theory of heat},
  author={Einstein, Albert and others},
  journal={Annalen der physik},
  volume={17},
  number={549-560},
  pages={208},
  year={1905}
}

@article{lee2020radial,
title = "The impact of realistic axonal shape on axon diameter estimation using diffusion {MRI}",
journal = "NeuroImage",
pages = "117228",
year = "2020",
issn = "1053-8119",
author = "Hong-Hsi Lee and Sune N. Jespersen and Els Fieremans and Dmitry S. Novikov",
}

@article{veraart2020highb,
  title={Noninvasive quantification of axon radii using diffusion {MRI}},
  author={Veraart, Jelle and Nunes, Daniel and Rudrapatna, Umesh and Fieremans, Els and Jones, Derek K and Novikov, Dmitry S and Shemesh, Noam},
  journal={eLife},
  volume={9},
  pages={e49855},
  year={2020},
  publisher={eLife Sciences Publications Limited}
}

@article{lee2020gray,
title = "In vivo observation and biophysical interpretation of time-dependent diffusion in human cortical gray matter",
journal = "NeuroImage",
pages = "117054",
year = "2020",
issn = "1053-8119",
author = "Hong-Hsi Lee and Antonios Papaioannou and Dmitry S. Novikov and Els Fieremans",
}

@article{lee2021rms,
  title={Realistic Microstructure Simulator (RMS): Monte Carlo simulations of diffusion in three-dimensional cell segmentations of microscopy images},
  author={Lee, Hong-Hsi and Fieremans, Els and Novikov, Dmitry S},
  journal={Journal of Neuroscience Methods},
  volume={350},
  pages={109018},
  year={2021},
  publisher={Elsevier}
}

@article{shapson2024petavoxel,
  title={A petavoxel fragment of human cerebral cortex reconstructed at nanoscale resolution},
  author={Shapson-Coe, Alexander and Januszewski, Micha{\l} and Berger, Daniel R and Pope, Art and Wu, Yuelong and Blakely, Tim and Schalek, Richard L and Li, Peter H and Wang, Shuohong and Maitin-Shepard, Jeremy and others},
  journal={Science},
  volume={384},
  number={6696},
  pages={eadk4858},
  year={2024},
  publisher={American Association for the Advancement of Science}
}

@article{kaden2016smt,
  title={Multi-compartment microscopic diffusion imaging},
  author={Kaden, Enrico and Kelm, Nathaniel D and Carson, Robert P and Does, Mark D and Alexander, Daniel C},
  journal={NeuroImage},
  volume={139},
  pages={346--359},
  year={2016},
  publisher={Elsevier}
}

@article{fan2020adm,
  title={Axon diameter index estimation independent of fiber orientation distribution using high-gradient diffusion MRI},
  author={Fan, Qiuyun and Nummenmaa, Aapo and Witzel, Thomas and Ohringer, Ned and Tian, Qiyuan and Setsompop, Kawin and Klawiter, Eric C and Rosen, Bruce R and Wald, Lawrence L and Huang, Susie Y},
  journal={Neuroimage},
  volume={222},
  pages={117197},
  year={2020},
  publisher={Elsevier}
}

@article{huang2021connectome,
  title={Connectome 2.0: Developing the next-generation ultra-high gradient strength human MRI scanner for bridging studies of the micro-, meso-and macro-connectome},
  author={Huang, Susie Y and Witzel, Thomas and Keil, Boris and Scholz, Alina and Davids, Mathias and Dietz, Peter and Rummert, Elmar and Ramb, Rebecca and Kirsch, John E and Yendiki, Anastasia and others},
  journal={NeuroImage},
  volume={243},
  pages={118530},
  year={2021},
  publisher={Elsevier}
}

@article{benjamini2016dpfg,
  title={White matter microstructure from nonparametric axon diameter distribution mapping},
  author={Benjamini, Dan and Komlosh, Michal E and Holtzclaw, Lynne A and Nevo, Uri and Basser, Peter J},
  journal={NeuroImage},
  volume={135},
  pages={333--344},
  year={2016},
  publisher={Elsevier}
}

@article{novikov2023exchange,
	Author = {Novikov, Dmitry S and Coronado-Leija, Ricardo and Fieremans, Els},
	Journal = {31st Annual Meeting of the International Society for Magnetic Resonance in Medicine, Toronto, Canada},
	Pages = {0684},
	Title = {Exchange between structurally-disordered compartments},
	Volume = {31},
	Year = {2023}
}

@article{donev2005packing,
  title={Neighbor list collision-driven molecular dynamics simulation for nonspherical hard particles. I. Algorithmic details},
  author={Donev, Aleksandar and Torquato, Salvatore and Stillinger, Frank H},
  journal={Journal of computational physics},
  volume={202},
  number={2},
  pages={737--764},
  year={2005},
  publisher={Elsevier}
}

@article{fieremans2011wmti,
  title={White matter characterization with diffusional kurtosis imaging},
  author={Fieremans, Els and Jensen, Jens H and Helpern, Joseph A},
  journal={Neuroimage},
  volume={58},
  number={1},
  pages={177--188},
  year={2011},
  publisher={Elsevier}
}

@article{chakwizira2023exchange,
  title={Diffusion MRI with pulsed and free gradient waveforms: effects of restricted diffusion and exchange},
  author={Chakwizira, Arthur and Westin, Carl-Fredrik and Brabec, Jan and Lasi{\v{c}}, Samo and Knutsson, Linda and Szczepankiewicz, Filip and Nilsson, Markus},
  journal={NMR in Biomedicine},
  volume={36},
  number={1},
  pages={e4827},
  year={2023},
  publisher={Wiley Online Library}
}

@article{dai2023og,
  title={Frequency-dependent diffusion kurtosis imaging in the human brain using an oscillating gradient spin echo sequence and a high-performance head-only gradient},
  author={Dai, Erpeng and Zhu, Ante and Yang, Grant K and Quah, Kristin and Tan, Ek T and Fiveland, Eric and Foo, Thomas KF and McNab, Jennifer A},
  journal={NeuroImage},
  volume={279},
  pages={120328},
  year={2023},
  publisher={Elsevier}
}

@article{xu2020breast,
  title={Magnetic resonance imaging of mean cell size in human breast tumors},
  author={Xu, Junzhong and Jiang, Xiaoyu and Li, Hua and Arlinghaus, Lori R and McKinley, Eliot T and Devan, Sean P and Hardy, Benjamin M and Xie, Jingping and Kang, Hakmook and Chakravarthy, A Bapsi and others},
  journal={Magnetic resonance in medicine},
  volume={83},
  number={6},
  pages={2002--2014},
  year={2020},
  publisher={Wiley Online Library}
}

@article{feinberg2023impulse,
  title={Next-generation MRI scanner designed for ultra-high-resolution human brain imaging at 7 Tesla},
  author={Feinberg, David A and Beckett, Alexander JS and Vu, An T and Stockmann, Jason and Huber, Laurentius and Ma, Samantha and Ahn, Sinyeob and Setsompop, Kawin and Cao, Xiaozhi and Park, Suhyung and others},
  journal={Nature methods},
  volume={20},
  number={12},
  pages={2048--2057},
  year={2023},
  publisher={Nature Publishing Group US New York}
}

@article{foo2020magnus,
  title={Highly efficient head-only magnetic field insert gradient coil for achieving simultaneous high gradient amplitude and slew rate at 3.0 T (MAGNUS) for brain microstructure imaging},
  author={Foo, Thomas KF and Tan, Ek Tsoon and Vermilyea, Mark E and Hua, Yihe and Fiveland, Eric W and Piel, Joseph E and Park, Keith and Ricci, Justin and Thompson, Paul S and Graziani, Dominic and others},
  journal={Magnetic resonance in medicine},
  volume={83},
  number={6},
  pages={2356--2369},
  year={2020},
  publisher={Wiley Online Library}
}

@article{palombo2020sandi,
  title={SANDI: A compartment-based model for non-invasive apparent soma and neurite imaging by diffusion MRI},
  author={Palombo, Marco and Ianus, Andrada and Guerreri, Michele and Nunes, Daniel and Alexander, Daniel C and Shemesh, Noam and Zhang, Hui},
  journal={Neuroimage},
  volume={215},
  pages={116835},
  year={2020},
  publisher={Elsevier}
}

@article{lee2024sandi,
  title={Age-related alterations in human cortical microstructure across the lifespan: Insights from high-gradient diffusion MRI},
  author={Lee, Hansol and Lee, Hong-Hsi and Ma, Yixin and Eskandarian, Laleh and Gaudet, Kyla and Tian, Qiyuan and Krijnen, Eva A and Russo, Andrew W and Salat, David H and Klawiter, Eric C and others},
  journal={Aging Cell},
  volume={23},
  number={11},
  pages={e14267},
  year={2024},
  publisher={Wiley Online Library}
}

@article{schiavi2023sandi,
  title={Mapping tissue microstructure across the human brain on a clinical scanner with soma and neurite density image metrics},
  author={Schiavi, Simona and Palombo, Marco and Zac{\`a}, Domenico and Tazza, Francesco and Lapucci, Caterina and Castellan, Lucio and Costagli, Mauro and Inglese, Matilde},
  journal={Human Brain Mapping},
  volume={44},
  number={13},
  pages={4792--4811},
  year={2023},
  publisher={Wiley Online Library}
}

@article{gyori2021sandi,
  title={On the potential for mapping apparent neural soma density via a clinically viable diffusion MRI protocol},
  author={Gyori, Noemi G and Clark, Christopher A and Alexander, Daniel C and Kaden, Enrico},
  journal={Neuroimage},
  volume={239},
  pages={118303},
  year={2021},
  publisher={Elsevier}
}

@article{giordano2003emt,
  title={Effective medium theory for dispersions of dielectric ellipsoids},
  author={Giordano, Stefano},
  journal={Journal of electrostatics},
  volume={58},
  number={1-2},
  pages={59--76},
  year={2003},
  publisher={Elsevier}
}

@article{sen1981emt,
  title={A self-similar model for sedimentary rocks with application to the dielectric constant of fused glass beads},
  author={Sen, PN and Scala, C and Cohen, MH},
  journal={Geophysics},
  volume={46},
  number={5},
  pages={781--795},
  year={1981},
  publisher={Society of Exploration Geophysicists}
}

@article{novikov2012emt,
  title={Relating extracellular diffusivity to cell size distribution and packing density as applied to white matter},
  author={Novikov, Dmitry S and Fieremans, Els},
  journal={Proceedings of ISMRM},
  volume={20},
  pages={1829},
  year={2012}
}

@article{latour1994time,
  title={Time-dependent diffusion of water in a biological model system.},
  author={Latour, Lawrence L and Svoboda, Karel and Mitra, Partha P and Sotak, Christopher H},
  journal={Proceedings of the National Academy of Sciences},
  volume={91},
  number={4},
  pages={1229--1233},
  year={1994}
}

@article{tomadakis1993transport,
  title={Transport properties of random arrays of freely overlapping cylinders with various orientation distributions},
  author={Tomadakis, Manolis M and Sotirchos, Stratis V},
  journal={The Journal of chemical physics},
  volume={98},
  number={1},
  pages={616--626},
  year={1993},
  publisher={American Institute of Physics}
}

@article{pallotto2015ecs,
  title={Extracellular space preservation aids the connectomic analysis of neural circuits},
  author={Pallotto, Marta and Watkins, Paul V and Fubara, Boma and Singer, Joshua H and Briggman, Kevin L},
  journal={Elife},
  volume={4},
  pages={e08206},
  year={2015},
  publisher={eLife Sciences Publications, Ltd}
}

@article{jan2010branching,
  title={Branching out: mechanisms of dendritic arborization},
  author={Jan, Yuh-Nung and Jan, Lily Yeh},
  journal={Nature Reviews Neuroscience},
  volume={11},
  number={5},
  pages={316--328},
  year={2010},
  publisher={Nature Publishing Group UK London}
}

@article{olesen2022smex,
  title={Diffusion time dependence, power-law scaling, and exchange in gray matter},
  author={Olesen, Jonas L and {\O}stergaard, Leif and Shemesh, Noam and Jespersen, Sune N},
  journal={NeuroImage},
  volume={251},
  pages={118976},
  year={2022},
  publisher={Elsevier}
}

@article{chan2025nexi,
  title={In vivo human neurite exchange time imaging at 500 mT/m diffusion gradients},
  author={Chan, Kwok-Shing and Ma, Yixin and Lee, Hansol and Marques, Jos{\'e} P and Olesen, Jonas L and Coelho, Santiago and Novikov, Dmitry S and Jespersen, Sune N and Huang, Susie Y and Lee, Hong-Hsi},
  journal={Imaging Neuroscience},
  volume={3},
  pages={imag\_a\_00544},
  year={2025},
  publisher={MIT Press 255 Main Street, 9th Floor, Cambridge, Massachusetts 02142, USA~…}
}

@article{jelescu2022nexi,
  title={Neurite Exchange Imaging (NEXI): A minimal model of diffusion in gray matter with inter-compartment water exchange},
  author={Jelescu, Ileana O and de Skowronski, Alexandre and Geffroy, Fran{\c{c}}oise and Palombo, Marco and Novikov, Dmitry S},
  journal={NeuroImage},
  volume={256},
  pages={119277},
  year={2022},
  publisher={Elsevier}
}

@article{uhl2024nexi,
  title={Quantifying human gray matter microstructure using neurite exchange imaging (NEXI) and 300 mT/m gradients},
  author={Uhl, Quentin and Pavan, Tommaso and Molendowska, Malwina and Jones, Derek K and Palombo, Marco and Jelescu, Ileana Ozana},
  journal={Imaging Neuroscience},
  volume={2},
  pages={1--19},
  year={2024},
  publisher={MIT Press One Broadway, 12th Floor, Cambridge, Massachusetts 02142, USA~…}
}

@article{coleman1969random,
  title={Random paths through convex bodies},
  author={Coleman, Rodney},
  journal={Journal of Applied Probability},
  volume={6},
  number={2},
  pages={430--441},
  year={1969},
  publisher={Cambridge University Press}
}

@book{chiu2013stochastic,
  title={Stochastic geometry and its applications},
  author={Chiu, Sung Nok and Stoyan, Dietrich and Kendall, Wilfrid S and Mecke, Joseph},
  year={2013},
  chapter={3},
  publisher={John Wiley \& Sons}
}

@article{ianucs2022sandimouse,
  title={Soma and Neurite Density MRI (SANDI) of the in-vivo mouse brain and comparison with the Allen Brain Atlas},
  author={Ianu{\c{s}}, Andrada and Carvalho, Joana and Fernandes, Francisca F and Cruz, Renata and Chavarrias, Cristina and Palombo, Marco and Shemesh, Noam},
  journal={NeuroImage},
  volume={254},
  pages={119135},
  year={2022},
  publisher={Elsevier}
}

@article{li2021cellad,
  title={Deep learning based neuronal soma detection and counting for Alzheimer's disease analysis},
  author={Li, Qiufu and Zhang, Yu and Liang, Hanbang and Gong, Hui and Jiang, Liang and Liu, Qiong and Shen, Linlin},
  journal={Computer Methods and Programs in Biomedicine},
  volume={203},
  pages={106023},
  year={2021},
  publisher={Elsevier}
}

@article{andrade2013cellad,
  title={Cell number changes in Alzheimer’s disease relate to dementia, not to plaques and tangles},
  author={Andrade-Moraes, Carlos Humberto and Oliveira-Pinto, Ana V and Castro-Fonseca, Emily and da Silva, Camila G and Guimaraes, Daniel M and Szczupak, Diego and Parente-Bruno, Danielle R and Carvalho, Ludmila RB and Polichiso, L{\'\i}via and Gomes, Bruna V and others},
  journal={Brain},
  volume={136},
  number={12},
  pages={3738--3752},
  year={2013},
  publisher={Oxford University Press}
}

@article{scott1992amygdala,
  title={Amygdala cell loss and atrophy in Alzheimer's disease},
  author={Scott, Samuel A and Sparks, D Larry and Scheff, Stephen W and Dekosky, Steven T and Knox, Craig A},
  journal={Annals of Neurology: Official Journal of the American Neurological Association and the Child Neurology Society},
  volume={32},
  number={4},
  pages={555--563},
  year={1992},
  publisher={Wiley Online Library}
}

@book{maxwell1873em,
  title={A treatise on electricity and magnetism},
  author={Maxwell, James Clerk},
  volume={1},
  year={1873},
  publisher={Clarendon press}
}

@article{maxwell1904emt,
  title={XII. Colours in metal glasses and in metallic films},
  author={Maxwell-Garnett, J Cl},
  journal={Philosophical Transactions of the Royal Society of London. Series A, Containing Papers of a Mathematical or Physical Character},
  volume={203},
  number={359-371},
  pages={385--420},
  year={1904},
  publisher={The Royal Society}
}

@article{bruggeman1935emt,
  title={Berechnung verschiedener physikalischer Konstanten von heterogenen Substanzen. I. Dielektrizit{\"a}tskonstanten und Leitf{\"a}higkeiten der Mischk{\"o}rper aus isotropen Substanzen},
  author={Bruggeman, Von DAG},
  journal={Annalen der physik},
  volume={416},
  number={7},
  pages={636--664},
  year={1935},
  publisher={Wiley Online Library}
}

@article{hashin1962emt,
  title={A variational approach to the theory of the effective magnetic permeability of multiphase materials},
  author={Hashin, Zvi and Shtrikman, Shmuel},
  journal={Journal of applied Physics},
  volume={33},
  number={10},
  pages={3125--3131},
  year={1962},
  publisher={American Institute of Physics}
}

@article{landauer1978emt,
    author = {Landauer, Rolf},
    title = {Electrical conductivity in inhomogeneous media},
    journal = {AIP Conference Proceedings},
    volume = {40},
    number = {1},
    pages = {2-45},
    year = {1978},
    month = {04},
    issn = {0094-243X},
    doi = {10.1063/1.31150},
    url = {https://doi.org/10.1063/1.31150},
    eprint = {https://pubs.aip.org/aip/acp/article-pdf/40/1/2/11845932/2\_1\_online.pdf},
}

@article{ramos2025connectome2,
  title={Ultra-high gradient connectomics and microstructure MRI scanner for imaging of human brain circuits across scales},
  author={Ramos-Llord{\'e}n, Gabriel and Lee, Hong-Hsi and Davids, Mathias and Dietz, Peter and Krug, Andreas and Kirsch, John E and Mahmutovic, Mirsad and M{\"u}ller, Alina and Ma, Yixin and Lee, Hansol and others},
  journal={Nature biomedical engineering},
  pages={1--16},
  year={2025},
  publisher={Nature Publishing Group UK London}
}

@misc{torquato2002emt,
  title={Random heterogeneous materials: microstructure and macroscopic properties},
  author={Torquato, Salvatore},
  volume={16},
  year={2002},
  publisher={Springer}
}

@article{kinney2013extracellular,
  title={Extracellular sheets and tunnels modulate glutamate diffusion in hippocampal neuropil},
  author={Kinney, Justin P and Spacek, Josef and Bartol, Thomas M and Bajaj, Chandrajit L and Harris, Kristen M and Sejnowski, Terrence J},
  journal={Journal of Comparative Neurology},
  volume={521},
  number={2},
  pages={448--464},
  year={2013},
  publisher={Wiley Online Library}
}

@article{markov2014determination,
  title={Determination of electrical conductivity of double-porosity formations by using generalized differential effective medium approximation},
  author={Markov, M and Mousatov, A and Kazatchenko, E and Markova, I},
  journal={Journal of Applied Geophysics},
  volume={108},
  pages={104--109},
  year={2014},
  publisher={Elsevier}
}

@article{tian2022comprehensive,
  title={Comprehensive diffusion MRI dataset for in vivo human brain microstructure mapping using 300 mT/m gradients},
  author={Tian, Qiyuan and Fan, Qiuyun and Witzel, Thomas and Polackal, Maya N and Ohringer, Ned A and Ngamsombat, Chanon and Russo, Andrew W and Machado, Natalya and Brewer, Kristina and Wang, Fuyixue and others},
  journal={Scientific data},
  volume={9},
  number={1},
  pages={7},
  year={2022},
  publisher={Nature Publishing Group UK London}
}

@article{harms2018extending,
  title={Extending the Human Connectome Project across ages: Imaging protocols for the Lifespan Development and Aging projects},
  author={Harms, Michael P and Somerville, Leah H and Ances, Beau M and Andersson, Jesper and Barch, Deanna M and Bastiani, Matteo and Bookheimer, Susan Y and Brown, Timothy B and Buckner, Randy L and Burgess, Gregory C and others},
  journal={Neuroimage},
  volume={183},
  pages={972--984},
  year={2018},
  publisher={Elsevier}
}

@article{bookheimer2019lifespan,
  title={The lifespan human connectome project in aging: an overview},
  author={Bookheimer, Susan Y and Salat, David H and Terpstra, Melissa and Ances, Beau M and Barch, Deanna M and Buckner, Randy L and Burgess, Gregory C and Curtiss, Sandra W and Diaz-Santos, Mirella and Elam, Jennifer Stine and others},
  journal={Neuroimage},
  volume={185},
  pages={335--348},
  year={2019},
  publisher={Elsevier}
}

@article{abdollahzadeh2021deepacson,
  title={DeepACSON automated segmentation of white matter in 3D electron microscopy},
  author={Abdollahzadeh, Ali and Belevich, Ilya and Jokitalo, Eija and Sierra, Alejandra and Tohka, Jussi},
  journal={Communications biology},
  volume={4},
  number={1},
  pages={179},
  year={2021},
  publisher={Nature Publishing Group UK London}
}

@article{wang2019echo,
  title={Echo planar time-resolved imaging (EPTI)},
  author={Wang, Fuyixue and Dong, Zijing and Reese, Timothy G and Bilgic, Berkin and Katherine Manhard, Mary and Chen, Jingyuan and Polimeni, Jonathan R and Wald, Lawrence L and Setsompop, Kawin},
  journal={Magnetic resonance in medicine},
  volume={81},
  number={6},
  pages={3599--3615},
  year={2019},
  publisher={Wiley Online Library}
}

@article{weissberg1963effective,
  title={Effective diffusion coefficient in porous media},
  author={Weissberg, Harold L},
  journal={Journal of Applied Physics},
  volume={34},
  number={9},
  pages={2636--2639},
  year={1963},
  publisher={American Institute of Physics}
}

@article{berryman1980long,
  title={Long-wavelength propagation in composite elastic media I. Spherical inclusions},
  author={Berryman, James G},
  journal={The Journal of the Acoustical Society of America},
  volume={68},
  number={6},
  pages={1809--1819},
  year={1980},
  publisher={Acoustical Society of America}
}

@article{nagelhus2013physiological,
  title={Physiological roles of aquaporin-4 in brain},
  author={Nagelhus, Erlend A and Ottersen, Ole P},
  journal={Physiological reviews},
  volume={93},
  number={4},
  pages={1543--1562},
  year={2013},
  publisher={American Physiological Society Bethesda, MD}
}

@article{lee2024effects,
  title={The effects of axonal beading and undulation on axonal diameter estimation from diffusion MRI: insights from simulations in human axons segmented from three-dimensional electron microscopy},
  author={Lee, Hong-Hsi and Tian, Qiyuan and Sheft, Maxina and Coronado-Leija, Ricardo and Ramos-Llorden, Gabriel and Abdollahzadeh, Ali and Fieremans, Els and Novikov, Dmitry S and Huang, Susie Y},
  journal={NMR in Biomedicine},
  volume={37},
  number={4},
  pages={e5087},
  year={2024},
  publisher={Wiley Online Library}
}

@article{abdollahzadeh2025scattering,
  title={Scattering approach to diffusion quantifies axonal damage in brain injury},
  author={Abdollahzadeh, Ali and Coronado-Leija, Ricardo and Lee, Hong-Hsi and Sierra, Alejandra and Fieremans, Els and Novikov, Dmitry S},
  journal={Nature Communications},
  volume={16},
  number={1},
  pages={9808},
  year={2025},
  publisher={Nature Publishing Group UK London}
}

@article{nguyen2026caterpillar,
  title={CATERPillar: a flexible framework for generating white matter numerical substrates with incorporated glial cells},
  author={Nguyen-Duc, Jasmine and Brammerloh, Malte and Cherchali, Melina and De Riedmatten, In{\`e}s and P{\'e}rot, Jean-Baptiste and Rafael-Pati{\~n}o, Jonathan and Jelescu, Ileana O},
  journal={Medical Image Analysis},
  pages={103946},
  year={2026},
  publisher={Elsevier}
}

@inproceedings{coronado2022time,
  title={Time-dependent diffusion and kurtosis in the extra-axonal space from 3d electron microscopy substrates of injured rat brain white matter},
  author={Coronado-Leija, Ricardo and Lee, Hong-Hsi and Abdollahzadeh, Ali and Tohka, Jussi and Sierra, Alejandra and Fieremans, Els and Novikov, D},
  booktitle={Joint Annual Meeting ISMRM-ESMRMB ISMRT 31st Annual Meeting, ISMRM-ESMRMB (ISMRM, 2022)},
  year={2022}
}

@article{van2014vivo,
  title={In vivo investigation of restricted diffusion in the human brain with optimized oscillating diffusion gradient encoding},
  author={Van, Anh T and Holdsworth, Samantha J and Bammer, Roland},
  journal={Magnetic resonance in medicine},
  volume={71},
  number={1},
  pages={83--94},
  year={2014},
  publisher={Wiley Online Library}
}

@article{sung2026complementary,
  title={Complementary Sensitivity of Fixed-Time and Fixed-Oscillation Regimes to Exchange and Structural Disorder in the Human Brain Revealed Using Oscillating-Gradient Diffusion MRI With Ultra-Strong Gradients},
  author={Sung, Dongsuk and Chan, Kwok-Shing and Gerold, Julianna and Zhong, Wen and Zheng, Jialan and Tian, Qiyuan and Guo, Hua and Huang, Susie Y and Lee, Hong-Hsi},
  journal={Magnetic Resonance in Medicine},
  volume={95},
  number={6},
  pages={3429--3444},
  year={2026},
  publisher={Wiley Online Library}
}

@article{zhu2025engineering,
  title={Engineering clinical translation of OGSE diffusion MRI},
  author={Zhu, Ante and Michael, Eric S and Li, Hua and Sprenger, Tim and Hua, Yihe and Lee, Seung-Kyun and Yeo, Desmond Teck Beng and McNab, Jennifer A and Hennel, Franciszek and Fieremans, Els and others},
  journal={Magnetic resonance in medicine},
  volume={94},
  number={3},
  pages={913--936},
  year={2025},
  publisher={Wiley Online Library}
}


\clearpage
\onecolumngrid
\setcounter{page}{1}
\setcounter{section}{0}
\setcounter{subsection}{0}
\renewcommand{\thefigure}{S\arabic{figure}}
\renewcommand{\thesection}{S\arabic{section}}
\setcounter{figure}{0}

\section*{Supplementary Materials}

\section{Estimation of cylinder number in overlapping configurations}
\label{sec:app}
Consider a medium composed of $N_c$ randomly positioned, randomly oriented thin cylinders of identical thickness. Cylinders may overlap with each other, and the overlapping volume is counted only once when estimating the volume fraction. We derive the relationship between $N_c$ and a target intra-cylindrical volume fraction $f_c$. Furthermore, we generalize this relationship to a medium composed not only of randomly oriented cylinders but also of randomly packed spheres with a volume fraction $f_s$.

\subsection{Randomly oriented cylinders generated via the homogeneous Poisson line process}

To model a distribution of randomly oriented and positioned cylinders in a three-dimensional domain, we adopt the homogeneous Poisson line process (mean free path randomness)\cite{coleman1969random} to generate infinite lines intersecting a cube of side length $L$. 
Each line is defined by a random orientation and a position sampled uniformly over the set of lines intersecting the cube. 
For each line, a cylinder of diameter $2r_c$ and length of the secant line inside the cube is retained. 
This approach ensures that the resulting cylinder geometry is statistically isotropic and homogeneous, consistent with properties of a three-dimensional Poisson line process.

The average length $\langle l \rangle$ of a line segment generated by this process within the cube is known from integral geometry \cite{coleman1969random}:
\begin{equation} \notag
    \langle l \rangle = \tfrac{2}{3} L.
\end{equation}
Hence, the average volume of a single cylinder in radius $r_c$ is
\begin{equation} \notag
    \langle v_c \rangle = \pi r_c^2 \cdot \langle l \rangle\,,
\end{equation}
given by Eq. (\ref{eq:v-cyl}).
To achieve a desired expected intra-cylindrical volume fraction $f_c$, we model the placement of cylinders as a Poisson process with intensity $N_c$, the expected number of cylinders in the cube. Since the cylinders can partially overlap, the net occupied volume is not exactly $N_c \cdot \langle v_c \rangle$, but instead follows the probabilistic coverage model \cite{chiu2013stochastic}:
\begin{equation} \notag
    f_c = 1 - \exp\left(-\frac{N_c \langle v_c \rangle }{L^3} \right).
\end{equation}
This equation accounts for the fact that overlapping volumes are only counted once, and is analogous to the well-known coverage formula in stochastic geometry.
Solving for $N_c$, we obtain
\begin{equation} \label{eq:Nc-cyl-only}
    N_c = -\frac{L^3}{\langle v_c \rangle} \cdot \ln(1 - f_c).
\end{equation}

\subsection{Correction for the Presence of Spheres}
If randomly packed non-overlapping spheres are also present in the cube, occupying a volume fraction $f_s$, they effectively reduce the available space in which cylinders can reside. While cylinders are still allowed to overlap with spheres, and that overlapping volume is counted toward the spheres, the available volume for cylinders must be adjusted to ensure the correct intra-cylindrical volume fraction achieved in the remaining space.

To compensate for this, we apply a volume exclusion correction to the target volume fraction $f_c$. The corrected formula becomes
\begin{equation} \notag
    f_c \to \frac{f_c}{1-f_s}\,.
\end{equation}
This correction arises from a mean-field assumption that cylinders are only occupying the volume not already filled by spheres.
Substituting this correction into Eq. (\ref{eq:Nc-cyl-only}) yields the adjusted number of cylinders needed in Eq. (\ref{eq:N-cyl}).
This correction ensures that the intra-cylindrical volume fraction is computed relative to the volume not already occupied by spheres, while still allowing geometrical overlaps between cylinders and spheres in the simulation.

\clearpage

\section{Oscillating-gradient waveform}
\label{sec:og-waveform}

\begin{figure*}[b!]
\centering
\includegraphics[width=0.675\textwidth]{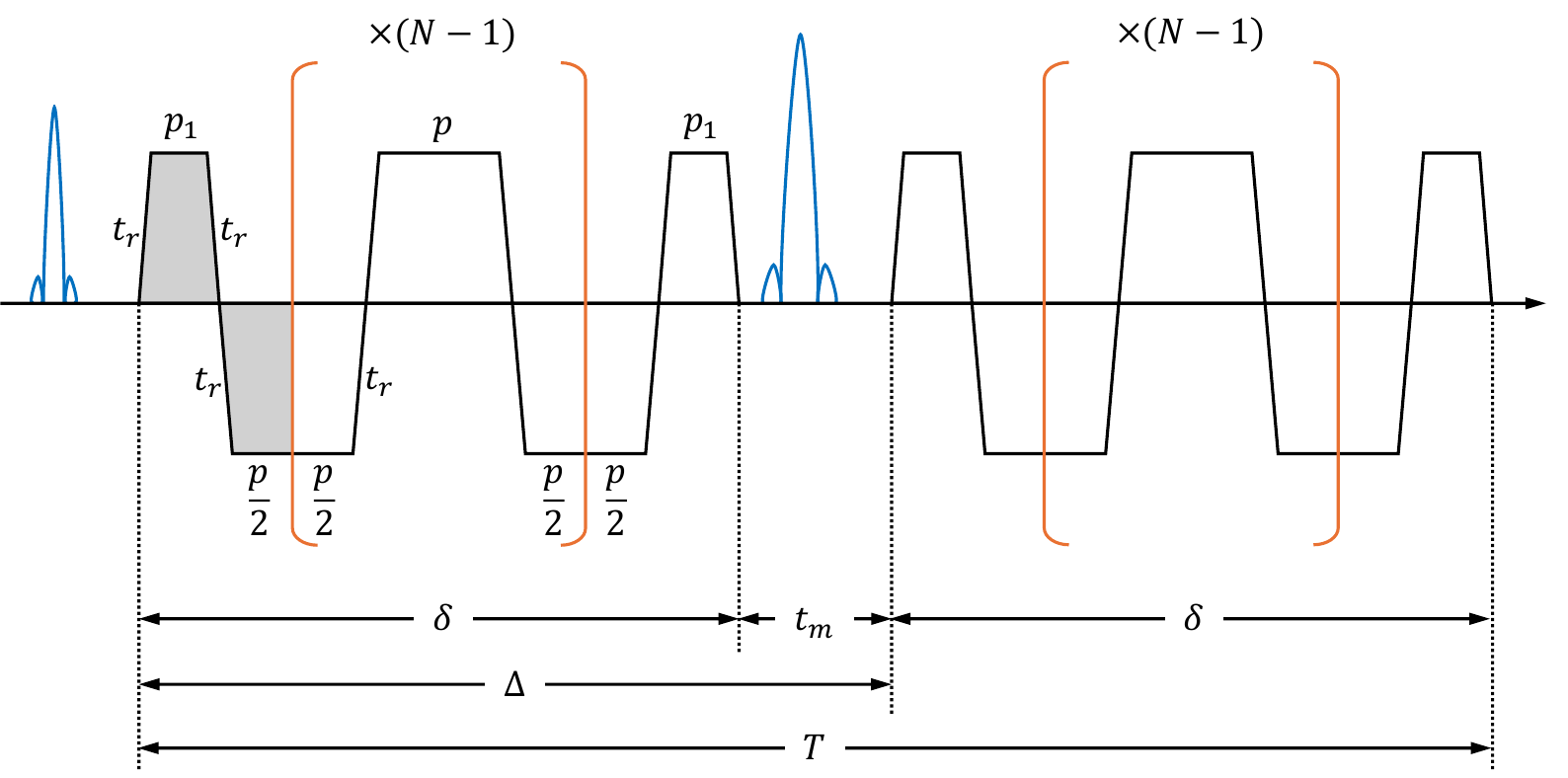}
\caption[]{
\textbf{} 
Oscillating-gradient waveform used in diffusion simulations.
The trapezoidal-cosine waveform can be parametrized by four quantities: the inter-waveform interval $\Delta$, waveform duration $\delta$, number of oscillations $N$, and rise time $t_r$.
Alternatively, it can be equivalently described by the total waveform time $T$, mixing time $t_m$, $N$, and $t_r$. 
The waveform amplitude is scaled by the Larmor gradient strength $g$.
Additional parameters, such as the first plateau time $p_1$ and the second plateau time $p$, are then determined accordingly.
Here, the actual gradient waveform applied on the scanner is shown, without accounting for the refocusing effect of the 180$^\circ$ pulse.
} 
\label{fig:og-waveform}
\end{figure*}

The oscillating-gradient waveform is parametrized by four sequence parameters (Figure~\ref{fig:og-waveform}): the inter-waveform interval $\Delta$, waveform duration $\delta$, number of oscillations $N$ on each side of the refocusing pulse, and the rise time (ramp time) $t_r$. 
Alternatively, it can be equivalently described by the total waveform duration $T$, mixing time $t_m$, $N$, and $t_r$, with the simple relations
\begin{equation*}
    T = \Delta + \delta\,,\quad t_m=\Delta-\delta\,.
\end{equation*}
The rise time is typically determined by the maximum gradient strength $G_\text{max}$ and the slew rate $SR$,
\begin{equation*}
    t_r\simeq G_\text{max}/SR\,,
\end{equation*}
where $G_\text{max}$ is in units of mT/m, and $SR$ is in units of T/m/s (equivalently, mT/m/ms).
The mixing time $t_m$ plays a critical role in shaping the power-spectrum of the waveform, which ideally exhibits a dominant main lobe with minimal, symmetrically distributed side lobes. 
Its optimal value can be estimated using approximate solutions based on the truncated cosine waveform \cite{van2014vivo}, or determined more precisely via numerical optimization for the trapezoidal-cosine waveform \cite{sung2026complementary}.
Additionally, the polarity of the second waveform after the 180$^\circ$ pulse can be inverted to further improve the spectral profile \cite{van2014vivo,dai2023og}.

Additional sequence parameters, such as the first plateau time $p_1$ and the second plateau time $p$, are then determined accordingly \cite{van2014vivo,baron2014ogse,xu2020breast},
\begin{equation*}
    p = 2p_1+t_r\,,\quad p_1 = \frac{\delta - t_r(6N+1)}{4N}\,.
\end{equation*}
Here, the first relation enforces the echo condition, $\int G(t')\,\mathrm{d}t'=0$, while the second follows from expressing $\delta$ in terms of $p_1$, $p$, $t_r$, and $N$. 
The conventional oscillating-gradient waveform (Figure~\ref{fig:og-waveform}) admits an analytical expression for the b-value \cite{van2014vivo,baron2014ogse,xu2020breast}
\begin{equation*}
    b=g^2 \cdot b_\text{unit}\,,\quad
    b_\text{unit} = \frac{91}{15}N t_r^3 + \frac{8}{3}Np_1^3 + \frac{1}{30} t_r^3 + 12 N t_r p_1^2 + \frac{46}{3} N t_r^2 p_1\,,
\end{equation*}
where $g=\gamma G$ is the Larmor gradient amplitude of the waveform.

\clearpage

\begin{figure}[th!]
\centering
\includegraphics[width=0.44\textwidth]{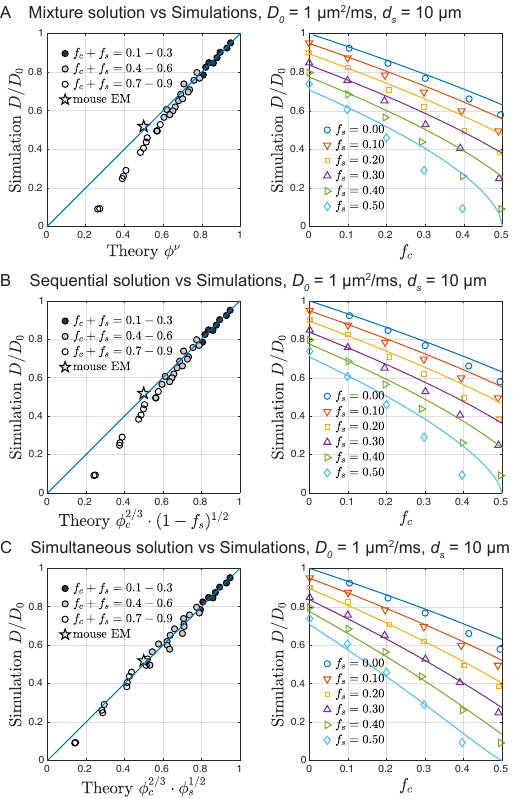}
\caption[]{\textbf{} Comparison of \textbf{A.} mixture solution (\ref{eq:sigma-emt-diff})--(\ref{nu}), \textbf{B.} sequential (\ref{eq:tortuosity-sequential}) and \textbf{C.} simultaneous coarse-graining solutions (\ref{eq:tortuosity-simultaneous}) for extra-cellular tortuosity with simulation results of intrinsic diffusivity $D_0=$ 1 \textmu m\textsuperscript{2}/ms in a medium composed of spheres of $d_s=$ 10 \textmu m in diameter and cylinders of 1 \textmu m in diameter.
In left panels, simulation results in extra-cellular space of mouse SBEM are shown in pentagrams.
In right panels, simulation results are shown as data points, and theoretical predictions are shown as curves.} 
\label{fig:geometry-D1-ds10}
\end{figure}

\begin{figure}[th!]
\centering
\includegraphics[width=0.44\textwidth]{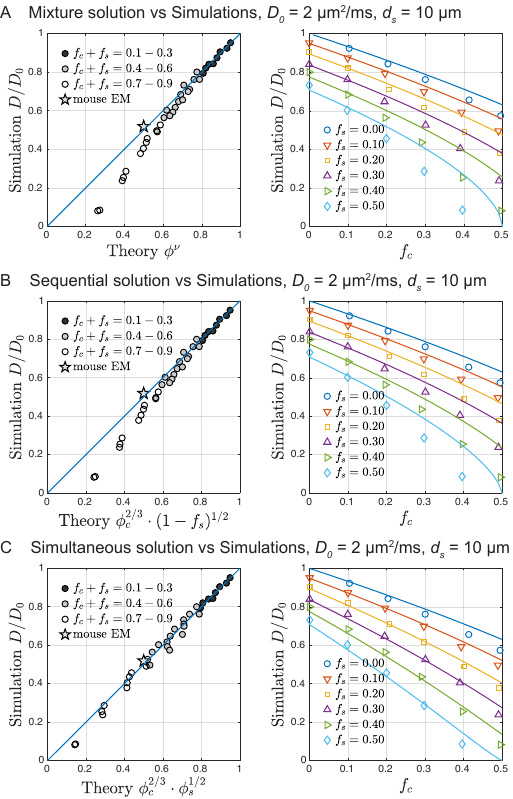}
\caption[]{\textbf{} Comparison of \textbf{A.} mixture solution (\ref{eq:sigma-emt-diff})--(\ref{nu}), \textbf{B.} sequential (\ref{eq:tortuosity-sequential}) and \textbf{C.} simultaneous coarse-graining solutions (\ref{eq:tortuosity-simultaneous}) for extra-cellular tortuosity with simulation results of intrinsic diffusivity $D_0=$ 2 \textmu m\textsuperscript{2}/ms in a medium composed of spheres of $d_s=$ 10 \textmu m in diameter and cylinders of 1 \textmu m in diameter.
In left panels, simulation results in extra-cellular space of mouse SBEM are shown in pentagrams.
In right panels, simulation results are shown as data points, and theoretical predictions are shown as curves.} 
\label{fig:geometry-D2-ds10}
\end{figure}

\begin{figure}[th!]
\centering
\includegraphics[width=0.44\textwidth]{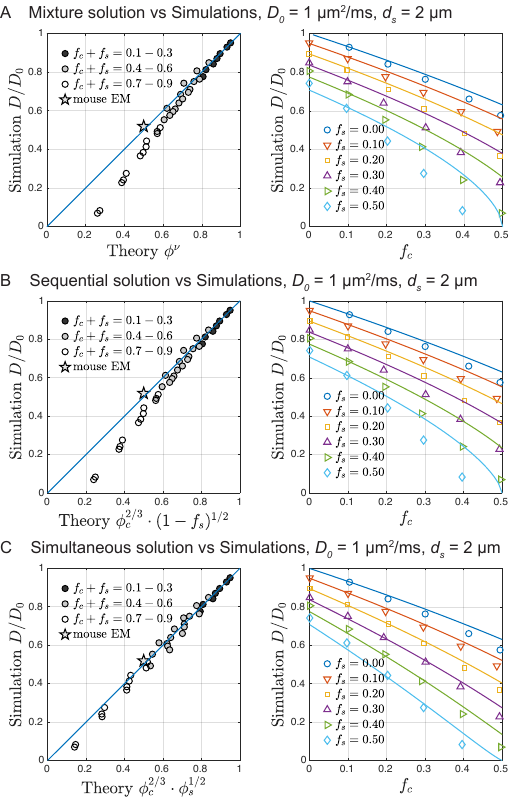}
\caption[]{\textbf{} Comparison of \textbf{A.} mixture solution (\ref{eq:sigma-emt-diff})--(\ref{nu}), \textbf{B.} sequential (\ref{eq:tortuosity-sequential}) and \textbf{C.} simultaneous coarse-graining solutions (\ref{eq:tortuosity-simultaneous}) for extra-cellular tortuosity with simulation results of intrinsic diffusivity $D_0=$ 1 \textmu m\textsuperscript{2}/ms in a medium composed of spheres of $d_s=$ 2 \textmu m in diameter and cylinders of 1 \textmu m in diameter.
In left panels, simulation results in extra-cellular space of mouse SBEM are shown in pentagrams.
In right panels, simulation results are shown as data points, and theoretical predictions are shown as curves.} 
\label{fig:geometry-D1-ds2}
\end{figure}

\begin{figure}[th!]
\centering
\includegraphics[width=0.44\textwidth]{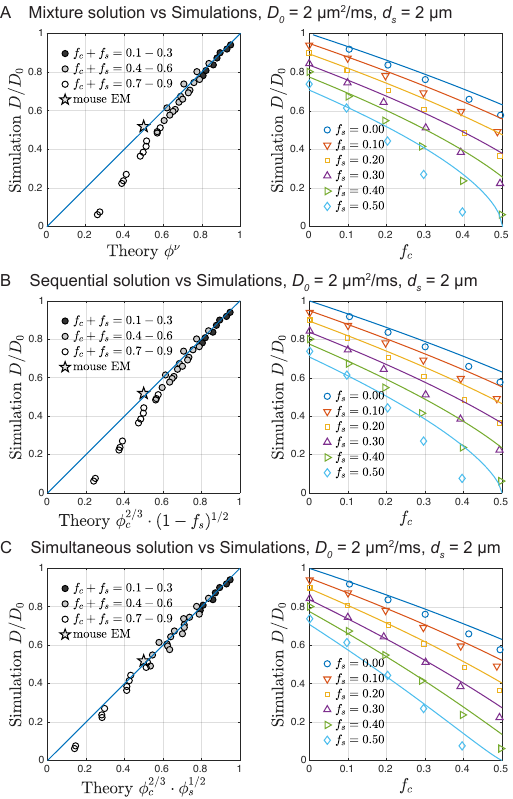}
\caption[]{\textbf{} Comparison of \textbf{A.} mixture solution (\ref{eq:sigma-emt-diff})--(\ref{nu}), \textbf{B.} sequential (\ref{eq:tortuosity-sequential}) and \textbf{C.} simultaneous coarse-graining solutions (\ref{eq:tortuosity-simultaneous}) for extra-cellular tortuosity with simulation results of intrinsic diffusivity $D_0=$ 2 \textmu m\textsuperscript{2}/ms in a medium composed of spheres of $d_s=$ 2 \textmu m in diameter and cylinders of 1 \textmu m in diameter.
In left panels, simulation results in extra-cellular space of mouse SBEM are shown in pentagrams.
In right panels, simulation results are shown as data points, and theoretical predictions are shown as curves.} 
\label{fig:geometry-D2-ds2}
\end{figure}

\begin{figure}[t!]
\centering
\includegraphics[width=0.45\textwidth]{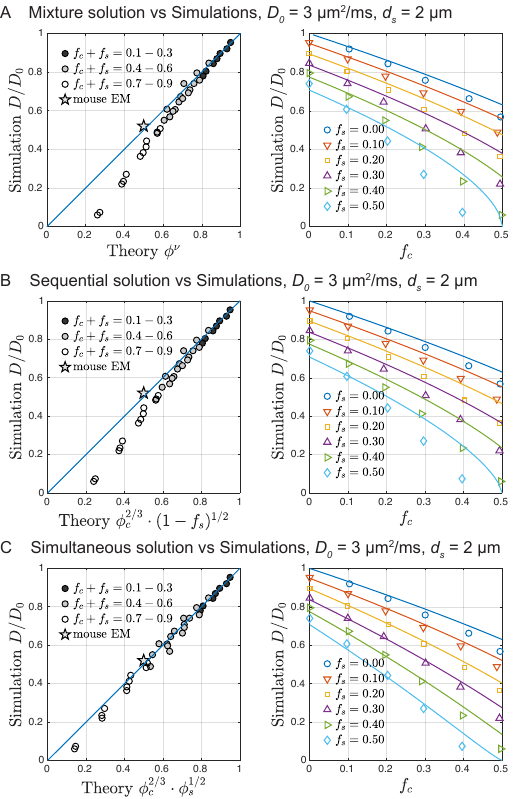}
\caption[]{\textbf{} Comparison of \textbf{A.} mixture solution (\ref{eq:sigma-emt-diff})--(\ref{nu}), \textbf{B.} sequential (\ref{eq:tortuosity-sequential}) and \textbf{C.} simultaneous coarse-graining solutions (\ref{eq:tortuosity-simultaneous}) for extra-cellular tortuosity with simulation results of intrinsic diffusivity $D_0=$ 3 \textmu m\textsuperscript{2}/ms in a medium composed of spheres of $d_s=$ 2 \textmu m in diameter and cylinders of 1 \textmu m in diameter.
In left panels, simulation results in extra-cellular space of mouse SBEM are shown in pentagrams.
In right panels, simulation results are shown as data points, and theoretical predictions are shown as curves.} 
\label{fig:geometry-D3-ds2}
\end{figure}

\clearpage

\begin{figure*}[t!]
\centering
\includegraphics[width=0.675\textwidth]{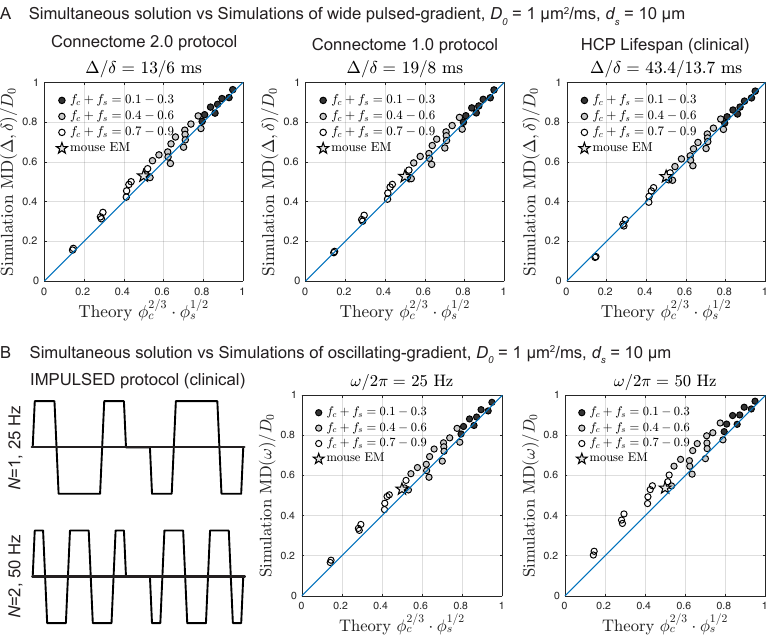}
\caption[]{
\textbf{} Comparison of simultaneous coarse-graining solution (\ref{eq:tortuosity-simultaneous}) for extra-cellular tortuosity with simulation results of \textbf{A.} wide pulsed-gradient and \textbf{B.} oscillating-gradient sequences.
We performed diffusion simulations with intrinsic diffusivity $D_0=$ 1 \textmu m\textsuperscript{2}/ms in a medium composed of spheres of $d_s=$ 10 \textmu m in diameter and cylinders of 1 \textmu m in diameter. 
For pulsed-gradient waveform, we simulated the time-dependent mean diffusivity $\text{MD}(\Delta,\delta)$ for diffusion protocols (inter-pulse interval $\Delta$, pulse width $\delta$) on three different scanners, such as Connectome 2.0 (500 mT/m) \cite{ramos2025connectome2}, Connectome 1.0 (300 mT/m) \cite{tian2022comprehensive}, and clinical scanners (80 mT/m) \cite{harms2018extending,bookheimer2019lifespan}.
For oscillating-gradient waveform, we simulated the frequency-dependent mean diffusivity $\text{MD}(\omega)$ for the IMPULSED protocol on clinical scanners (80 mT/m) \cite{xu2020breast}.
Here, we showed effective gradient waveforms of oscillating-gradient sequences, where the gradient polarity after the 180$^\circ$ pulse was inverted to account for the refocusing effect.
} 
\label{fig:pgse-D1-ds10}
\end{figure*}

\begin{figure*}[t!]
\centering
\includegraphics[width=0.675\textwidth]{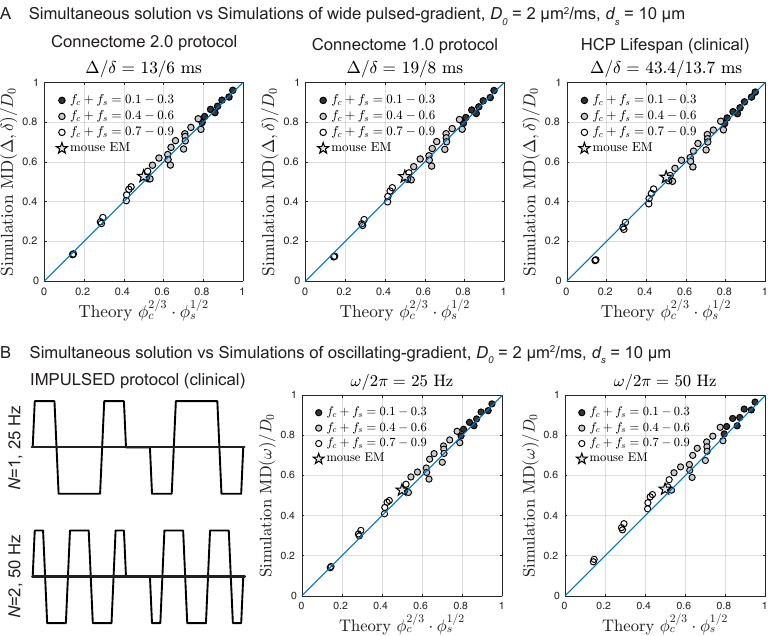}
\caption[]{
\textbf{} Comparison of simultaneous coarse-graining solution (\ref{eq:tortuosity-simultaneous}) for extra-cellular tortuosity with simulation results of \textbf{A.} wide pulsed-gradient and \textbf{B.} oscillating-gradient sequences.
We performed diffusion simulations with intrinsic diffusivity $D_0=$ 2 \textmu m\textsuperscript{2}/ms in a medium composed of spheres of $d_s=$ 10 \textmu m in diameter and cylinders of 1 \textmu m in diameter. 
For pulsed-gradient waveform, we simulated the time-dependent mean diffusivity $\text{MD}(\Delta,\delta)$ for diffusion protocols (inter-pulse interval $\Delta$, pulse width $\delta$) on three different scanners, such as Connectome 2.0 (500 mT/m) \cite{ramos2025connectome2}, Connectome 1.0 (300 mT/m) \cite{tian2022comprehensive}, and clinical scanners (80 mT/m) \cite{harms2018extending,bookheimer2019lifespan}.
For oscillating-gradient waveform, we simulated the frequency-dependent mean diffusivity $\text{MD}(\omega)$ for the IMPULSED protocol on clinical scanners (80 mT/m) \cite{xu2020breast}.
Here, we showed effective gradient waveforms of oscillating-gradient sequences, where the gradient polarity after the 180$^\circ$ pulse was inverted to account for the refocusing effect.
} 
\label{fig:pgse-D2-ds10}
\end{figure*}

\clearpage

\begin{figure*}[t!]
\centering
\includegraphics[width=0.675\textwidth]{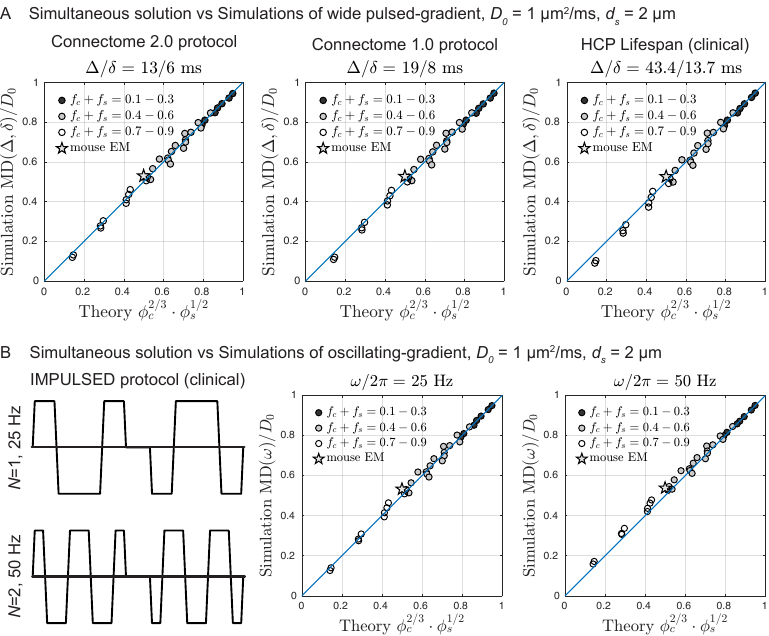}
\caption[]{\textbf{} 
\textbf{} Comparison of simultaneous coarse-graining solution (\ref{eq:tortuosity-simultaneous}) for extra-cellular tortuosity with simulation results of \textbf{A.} wide pulsed-gradient and \textbf{B.} oscillating-gradient sequences.
We performed diffusion simulations with intrinsic diffusivity $D_0=$ 1 \textmu m\textsuperscript{2}/ms in a medium composed of spheres of $d_s=$ 2 \textmu m in diameter and cylinders of 1 \textmu m in diameter. 
For pulsed-gradient waveform, we simulated the time-dependent mean diffusivity $\text{MD}(\Delta,\delta)$ for diffusion protocols (inter-pulse interval $\Delta$, pulse width $\delta$) on three different scanners, such as Connectome 2.0 (500 mT/m) \cite{ramos2025connectome2}, Connectome 1.0 (300 mT/m) \cite{tian2022comprehensive}, and clinical scanners (80 mT/m) \cite{harms2018extending,bookheimer2019lifespan}.
For oscillating-gradient waveform, we simulated the frequency-dependent mean diffusivity $\text{MD}(\omega)$ for the IMPULSED protocol on clinical scanners (80 mT/m) \cite{xu2020breast}.
Here, we showed effective gradient waveforms of oscillating-gradient sequences, where the gradient polarity after the 180$^\circ$ pulse was inverted to account for the refocusing effect.
} 
\label{fig:pgse-D1-ds2}
\end{figure*}

\begin{figure*}[t!]
\centering
\includegraphics[width=0.675\textwidth]{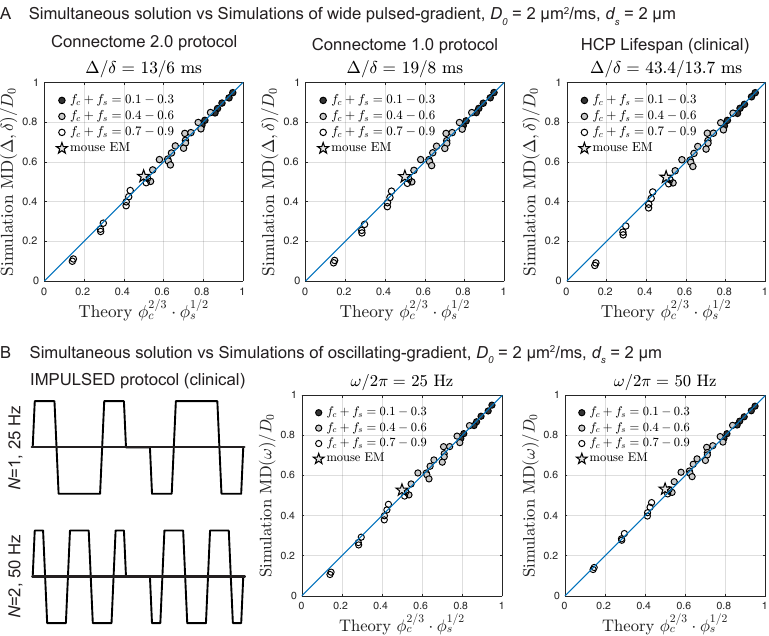}
\caption[]{
\textbf{} Comparison of simultaneous coarse-graining solution (\ref{eq:tortuosity-simultaneous}) for extra-cellular tortuosity with simulation results of \textbf{A.} wide pulsed-gradient and \textbf{B.} oscillating-gradient sequences.
We performed diffusion simulations with intrinsic diffusivity $D_0=$ 2 \textmu m\textsuperscript{2}/ms in a medium composed of spheres of $d_s=$ 2 \textmu m in diameter and cylinders of 1 \textmu m in diameter. 
For pulsed-gradient waveform, we simulated the time-dependent mean diffusivity $\text{MD}(\Delta,\delta)$ for diffusion protocols (inter-pulse interval $\Delta$, pulse width $\delta$) on three different scanners, such as Connectome 2.0 (500 mT/m) \cite{ramos2025connectome2}, Connectome 1.0 (300 mT/m) \cite{tian2022comprehensive}, and clinical scanners (80 mT/m) \cite{harms2018extending,bookheimer2019lifespan}.
For oscillating-gradient waveform, we simulated the frequency-dependent mean diffusivity $\text{MD}(\omega)$ for the IMPULSED protocol on clinical scanners (80 mT/m) \cite{xu2020breast}.
Here, we showed effective gradient waveforms of oscillating-gradient sequences, where the gradient polarity after the 180$^\circ$ pulse was inverted to account for the refocusing effect.
} 
\label{fig:pgse-D2-ds2}
\end{figure*}

\begin{figure*}[t!]
\centering
\includegraphics[width=0.675\textwidth]{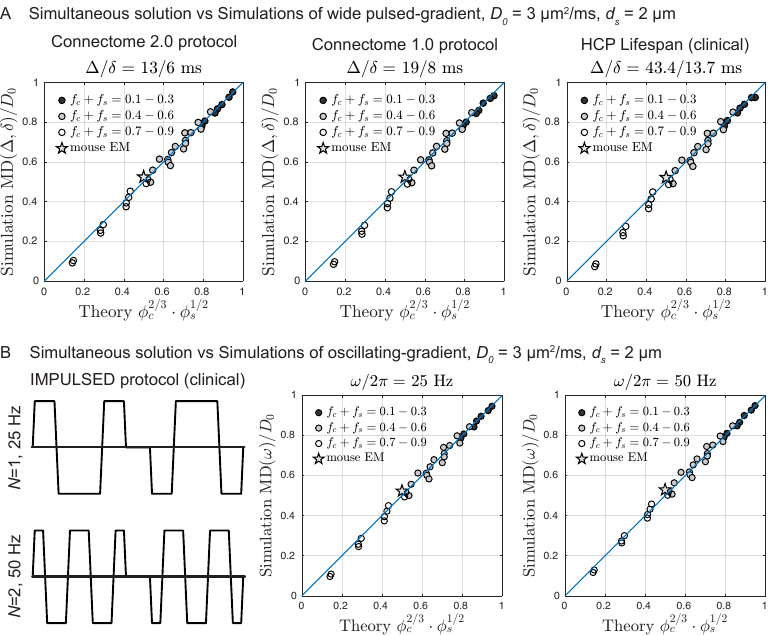}
\caption[]{
\textbf{} Comparison of simultaneous coarse-graining solution (\ref{eq:tortuosity-simultaneous}) for extra-cellular tortuosity with simulation results of \textbf{A.} wide pulsed-gradient and \textbf{B.} oscillating-gradient sequences.
We performed diffusion simulations with intrinsic diffusivity $D_0=$ 3 \textmu m\textsuperscript{2}/ms in a medium composed of spheres of $d_s=$ 2 \textmu m in diameter and cylinders of 1 \textmu m in diameter. 
For pulsed-gradient waveform, we simulated the time-dependent mean diffusivity $\text{MD}(\Delta,\delta)$ for diffusion protocols (inter-pulse interval $\Delta$, pulse width $\delta$) on three different scanners, such as Connectome 2.0 (500 mT/m) \cite{ramos2025connectome2}, Connectome 1.0 (300 mT/m) \cite{tian2022comprehensive}, and clinical scanners (80 mT/m) \cite{harms2018extending,bookheimer2019lifespan}.
For oscillating-gradient waveform, we simulated the frequency-dependent mean diffusivity $\text{MD}(\omega)$ for the IMPULSED protocol on clinical scanners (80 mT/m) \cite{xu2020breast}.
Here, we showed effective gradient waveforms of oscillating-gradient sequences, where the gradient polarity after the 180$^\circ$ pulse was inverted to account for the refocusing effect.
} 
\label{fig:pgse-D3-ds2}
\end{figure*}


\end{document}